\documentclass{aa}  

\usepackage{graphicx}
\usepackage{overpic}
\usepackage{xcolor}
\usepackage{natbib}
\bibpunct{(}{)}{;}{a}{}{,}
\usepackage{placeins}
\usepackage{txfonts}
\newcommand{\sub}[1]{_{\rm #1}}
\newcommand{\CII}{[C\,{\sc ii}]}   
\newcommand{\CI}{[C\,{\sc i}]}  
\newcommand{\HI}{H\,{\sc i}} 
\newcommand{\emm}[1]{\ensuremath{#1}}   
\newcommand{\emr}[1]{\emm{\mathrm{#1}}} 
\newcommand{\unit}[1]{\emm{\, \emr{#1}}}

\newcommand{\Kkms}{\unit{K\,km\,s^{-1}}}

\newcommand{\pscm}{\unit{cm^{-2}}}

\newcommand{\pccm}{\unit{cm^{-3}}}
\newcommand{\kms}{\unit{km\,s^{-1}}}

\newcommand{\changed}[1]{#1}

\begin{document}

   \title{Extended atomic carbon around molecular clouds}

   \author{V. Ossenkopf-Okada
          \inst{1}
          \and
          A. Karska\inst{2,3}
          \and
          M. Benedettini\inst{4}
          \and
          D. Colombo\inst{5}
          \and
          R. Simon\inst{1}
          }

   \institute{Universit\"at zu K\"oln, I. Physikalisches Institut, Z\"ulpicher Str. 77, 50937 K\"oln, Germany\\
    \email{ossk@ph1.uni-koeln.de}
    \and
    Max-Planck-Insitut für Radioastronomie, Auf dem Hügel 69, 53121 Bonn, Germany
    \and
    Institute of Astronomy, Faculty of Physics, Astronomy and Informatics, Nicolaus Copernicus University, Grudzi\k{a}dzka 5, 87-100 Toruń, Poland
    \and
    INAF - Istituto di Astrofisica e Planetologia Spaziali, Via Fosso del Cavaliere 100, I-00133 Roma, Italy
    \and
    Argelander-Institut für Astronomie, Universität Bonn, Auf dem H\"ugel 71, 53121 Bonn, Germany
 }

   \date{Received; accepted}

 
  \abstract
   {Models predict that atomic carbon occurs at the surface and in the process of the formation of molecular clouds, making its fine structure transitions a diagnostic of cloud formation.}
   {We study the distribution of atomic carbon in a small inconspicuous region towards the outer Galaxy that might be representative for a large fraction of the molecular gas of the Milky Way that is not directly affected by star formation.}
   {We observed a small strip of 5~arcminutes in the ``Forgotten Quadrant'', the third quadrant of the Milky Way, with the APEX telescope in the $^3P_1-^3P_0$ \CI{} transition of atomic carbon and the $J=2-1$ transition of the three most abundant CO isotopologues and compared their distribution with existing measurements of gas column density and of ionized carbon.}
   {The atomic carbon shows a very smooth distribution with the smallest gradient along the strip compared to the other lines. It is always brighter than $^{13}$CO and in one velocity-component even brighter than CO. In contrast to observations of many star-forming regions, the \CI{} emission seems to extend beyond the molecular gas, in line with the models of photon-dominated regions (PDRs). However, a standard PDR model fit to the observations fails because the models either predict more molecular gas, traced through C$^{18}$O, or more diffuse gas, traced through \CII{}, than observed. The carbon-budget in the gas phase does not add up to the same column seen through dust emission.}
   {To understand the \CI{} emission from galaxies it is necessary to get the full statistics for the quiescent gas outside of the star-forming regions that behaves significantly different from dense gas exposed to high ultraviolet fields.}


   \maketitle
%

\section{Introduction}

Every surface of a molecular cloud provides a transition from a diffuse atomic phase, where most molecules are dissociated by the ambient ultraviolet (UV) field, to a denser and colder phase of molecular hydrogen containing a variety of other molecules when shielded from the UV field. The transition is described in terms of photon-dominated or photo-dissociation regions \citep[PDRs,][]{HollenbachTielens1999, Wolfire2022}. PDR models predict the chemical and temperature structure of this transition and have been widely used to interpret observations of the interstellar medium at cloud surfaces \citep{Ossenkopf2007, Roellig2007, Wolfire2022}. 
The models predict a chemical layering where the carbon in the gas phase is predominantly singly ionized in the outer layers of the region, then turns neutral at intermediate depths where UV photons are sufficiently shielded by interstellar dust, and finally is seen mainly molecular in CO. As CO is self-shielding from the impinging UV radiation, the different CO isotopologues form at different depths according to their elemental abundance, $^{12}$CO forms closest to the cloud surface, $^{13}$CO somewhat deeper in and C$^{18}$O only deep in the clouds \citep[e.g.][]{Visser2009}.
As a consequence, we expect from the models that atomic carbon behaves as a surface tracer being sandwiched between ionized carbon and CO. 

In the global dynamical picture, atomic carbon is a valuable tracer of cloud mass accretion via streams of atomic and CO-dark molecular gas onto molecular clouds \citep{Wolfire2010,Glover2015}. Compared to ionized carbon, 
it suffers much less from line of sight confusion with other phases of ISM. A comparison of the distribution of atomic and molecular carbon can therefore quantify the amount of ongoing mass accretion onto clouds,  feeding future star-formation.

However, observations of atomic carbon through the \mbox{$^3P_1-^3P_0$} \CI{} transition at 492.2~GHz, for the rest of the paper abbreviated as \CI{} 1-0,  are still notoriously difficult due to the limited atmospheric transmission at that frequency. Therefore, many observations have been limited so far to bright sources with high densities and temperatures. 
\CI{} has been mapped in several active star-forming regions where line intensities up to some ten Kelvin were detected \citep[e.g.][]{Shimajiri2013, PerezBeaupuits2015}. Unfortunately, the observations often show a contradiction to the models. The carbon-chemistry models predict that atomic carbon should be present in relatively large abundances in a layer around the molecular material, but in many cases \CI{} is detected in the inside of the molecular medium, matching the distribution of $^{13}$CO \citep{Shimajiri2013,PerezBeaupuits2015,Lee2022} and being less abundant than predicted \citep{Gong2017, Roellig2013}. 
For the Orion A cloud, \citet{Arunachalam2023} and \citet{Labkhandifar2023} found the best correlation between \CI{} 1-0 and $^{13}$CO 2-1. The good correlation was confirmed for a sample of 133 massive ATLASGAL clumps in the inner Galaxy by \citet{Lee2022} but by including weaker sources they obtained a somewhat higher ratio between the two lines.
In contrast, \citet{Plume2000} found significantly more extended \CI{} emission in Orion A, however only at a resolution of about 3 arcminutes, compared to $^{13}$CO 1-0 data at much higher resolution. In general, the width of this PDR transition layer, where atomic carbon is abundant, is still largely unknown.  

The bright regions observed so far may be of secondary importance for the global carbon budget over large scales where most of the material is much more diffuse. 
The integrated emission over large areas or whole galaxies may be rather dominated by material at low and intermediate densities ($<~10^3$~cm$^{-3}$), partially CO-dark, not by material in the brightest dense star-forming cores \citep[see e.g.][]{Alaghband-Zadeh2013}. 
Some spot observations of atomic carbon from the diffuse medium were obtained in the UV towards bright background point sources \citep{Jenkins2011, Godard2024} detecting significant amounts of \CI{}, yet still no large-scale statistics are available.

A statistically complete answer to these questions is promised by the Galactic Ecology
large-scale \CI{} survey \citep[GEco,][]{Schilke2015} to be performed with the Fred Young Submillimeter Telescope (FYST) of the Cerro Chajnantor Atacama Telescope (CCAT) Observatory, starting early 2026. However, due to the largely unknown extent and intensity of the low-level \CI{} emission, preparatory observations covering some diffuse regions are needed to plan the GEco observations. 

   
Therefore, we performed a path-finder observation with the Atacama Pathfinder Experiment \citep[APEX,][]{Guesten2006} of a small field potentially representative for the extended diffuse medium around molecular clouds. It is a cut across a molecular cloud edge where the \CI{} emission should allow us to trace the material in the dynamic cloud boundary layer.

In Sect.~\ref{sect_2} we describe the observations and the data reduction. Sect.~\ref{sect_results} shows the spatial and velocity structure seen in the different tracers and Sect.~\ref{sect_ratios} compares their distribution for the individual spectral pixels. In Sect.~\ref{sect_models} we try to interpret the measurements in terms of models with increasing level of sophistication while Sect.~\ref{sect_discussion} discusses the outcomes, in particular the failures of all the models. We close in Sect.~\ref{sect_conclusions} with some conclusions that go beyond the scale of our small-field observations.
\section{Observations}
\label{sect_2}

\begin{figure}
   \centering
   \includegraphics[width=0.9\columnwidth]{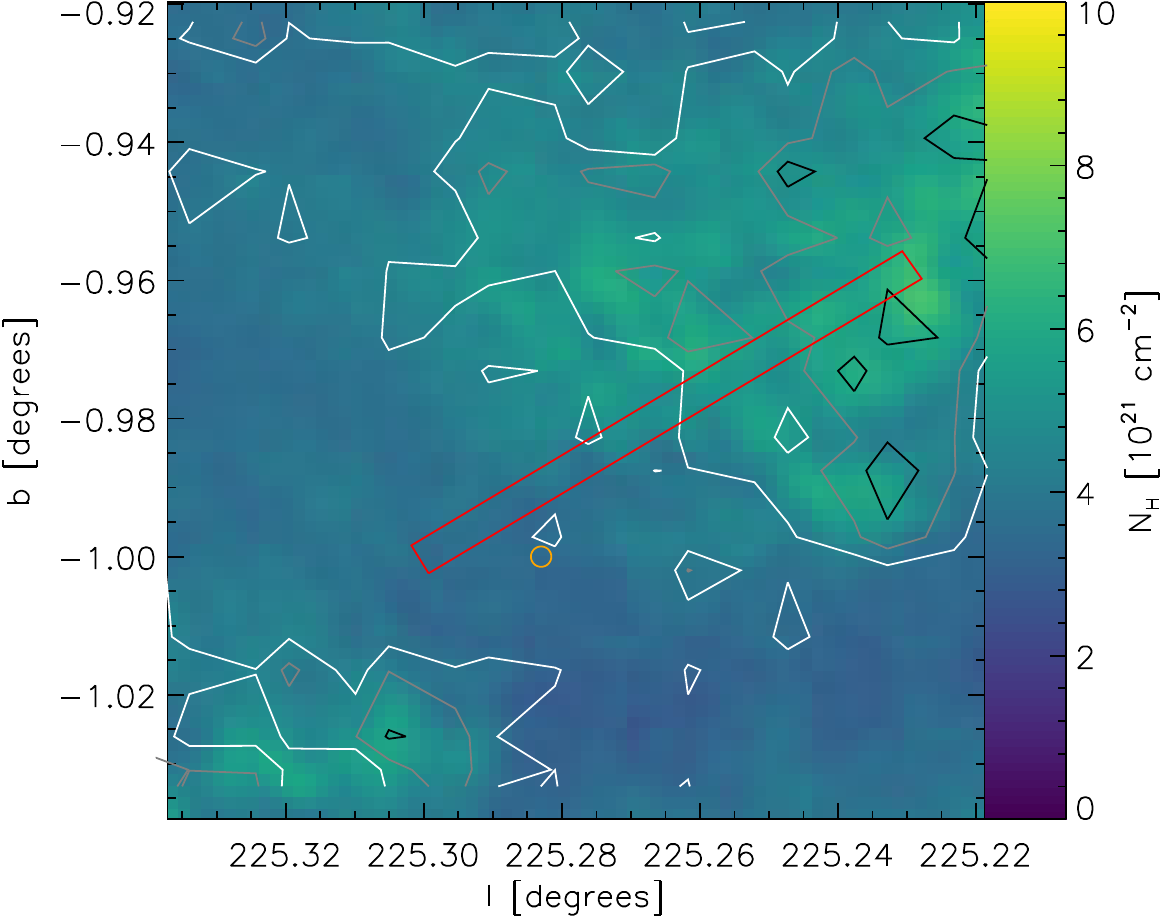}
      \caption{Known structure of the environment of the mapped strip. The colors give the total gas column density derived from the \textit{Herschel} SPIRE and PACS observations. The contours show the line integrated intensity of $^{13}$CO 1-0 from the Forgotten Quadrant Survey at the levels of 1.5 (white), 3.0 (gray), and 4.5 (black)~\Kkms{}. The mapped area is shown as a red rectangle. The \changed{orange} circle gives the beam and position of the existing single pointing \CII{} observation.
     }
         \label{fig_fqs}
\end{figure}

Because of the strong effect of the atmospheric transmission on the quality of the data, the APEX observations were done on a field that culminates almost in zenith at the telescope site minimizing the air mass. 
We observed a strip with a length of 5~arcminutes at the boundary of the FQS-MC224.955-0.671 molecular cloud with intermediate column densities and weak $^{13}$CO 1-0 (see Fig.~\ref{fig_fqs}). 
It is basically a one-dimensional cut across the molecular cloud edge, but to allow for convolution to the complementary data, it was set up as a narrow map with a width of 21~arcsec. 
The region was recently mapped in CO and $^{13}$CO by the 12~m antenna of the Arizona Radio Observatory in the frame of the ``Forgotten Quadrant Survey'' \citep[FQS,][]{Benedettini2020,Benedettini2021} and the cloud properties were analyzed by \citet{Dong2023}. Along our strip, the total column density of hydrogen changes from $N($H$)=3\,10^{-21}$~cm$^{-2}$ to $6\,10^{-21}$~cm$^{-2}$ \citep[$A_V=1.5\dots 3$, Herschel Hi-GAL,][]{Molinari2010}. $^{13}$CO 1-0 was not detected at the lower-column density end of the strip while it is clearly detected at the other end of the strip. All lines are narrow with a full-width-half-maximum (FWHM) of less than 3~\kms{}. The strip \changed{was planned to contain} a Herschel pointing in the GOTC+ survey measuring the \CII{} emission \changed{\citep[][at l,b=225.283,-1.0 deg]{Pineda2013} that would allow us to analyze the full carbon budget at this position. Unfortunately, the observations were carried out with a small offset (see Fig.~\ref{fig_fqs}) relative to the GOTC+ position and} the \CII{} observation only provided a tentative detection or an upper limit.  

The source was observed during the mornings of September 15, 21, and 22, 2024 at good weather conditions with a perceptible water vapor column of less than 1~mm.
On-the-Fly (OTF) observations were performed using the nFLASH460 receiver for the \CI{} line at 492~GHz and nFLASH230 running in parallel optimized for the C$^{18}$O 2-1 line at 220~GHz. The corresponding transition of the main CO isotopologue and of $^{13}$CO came as a side-product within the IF bandpass. No other lines were detected. 
Individual coverages used dump times of 0.5 seconds and a sampling of 4.2~arcsec corresponding to about 1/3 of the beam at 492~GHz. After a total observing time of 10.4~h the radiometric noise root-mean-square (RMS) was predicted to fall at 89~mK in antenna temperature for individual pointings in \CI{} at a velocity resolution of 0.5~\kms{}.
The observations provided a spatial resolution of 27~arcsec at 220-230~GHz and 13~arcsec at 492~GHz and a velocity resolution of 0.083~\kms{}.

\subsection{Data reduction}
\label{sect_observations}

All spectra were calibrated with the standard pipeline integrated in the APEX Control System \citep[APECS,][]{Muders2006}. Based on the main beam efficiencies measured on Mars and Jupiter in fall 2024, the spectra were calibrated from antenna temperatures to main beam temperatures, using a forward efficiency of 0.985 and main beam efficiencies at 230 and 490 GHz of 0.8 and 0.6, respectively. The efficiencies have an uncertainty of slightly less than 10\,\%. This should be taken into account as a systematic uncertainty of the data. No baseline subtraction apart from a constant offset, this means a zero-order baseline subtraction, was needed.

Using Gaussian kernels, the spectra were resampled to different spatial-spectral grids. For plotting, all results were resampled to a common grid in R.A.-Dec with a resolution of 29.7~arcsec, 10\,\% larger than the telescope resolution at 220~GHz and a spacing of 6.75~arcsec. For the quantitative analyzes in Sect.~\ref{sect_ratios} we resampled to a uniform grid with a sampling of 4.2~arcsec parallel and perpendicular to the strip direction and the same effective resolution of 29.7~arcsec. For the investigation of the spatial variations of the \CI{} line in Appx.~\ref{appx_highres} we resampled the data to a resolution of 14.3~arcsec keeping the spacing of 4.2~arcsec. To reduce the noise in the data, they were all spectrally resampled to a common velocity resolution of 0.25~\kms{}.

The investigated data cubes had a mean RMS noise level at the main beam temperature scale of 11~mK for C$^{18}$O 2-1 and $^{13}$CO 2-1, 12~mK for CO 2-1, and 47~mK for \CI{} at 29.7~arcsec resolution and 160~mK at the resolution of 14.3~arcsec.

As zero position of all data, we define the location of the existing GOTC+ pointing, $l=225.3, b=-1.0$ in Galactic coordinates or R.A.=07:10:39.8, Dec= -11:27:09 in J2000 equatorial coordinates.

\subsection{Complementary data}
\label{sect_complementary}

The total column density of the region based on dust emission was retrieved from the
HiGAL-PPMAP online database\footnote{\url{http://www.astro.cardiff.ac.uk/research/ViaLactea/}}, that translated all Herschel Hi-GAL data of the Galactic Plane \citep{Molinari2010} into dust temperatures and column densities per temperature bin using the PPMAP algorithm \citep{Marsh2015, Marsh2017}. The measured dust opacity is translated into gas column densities using a fixed opacity of 0.1 cm$^2/$g at 300~$\mu$m and the result is given in terms of the column density of hydrogen molecules. To avoid any assumption about the molecular state of hydrogen, we here use the more general definition $N\sub{H}=N(\mathrm{H})+2N(\mathrm{H}_2)$, effectively multiplying the PPMAP data by two. The maps have a spatial resolution of the reconstructed column density of 12~arcsec. 

CO 1-0 and $^{13}$CO 1-0 data were taken from the Forgotten Quadrant Survey \citep{Benedettini2020,Benedettini2021}. They have a spatial resolution of 55~arcsec and a spectral resolution of 0.3~\kms{}. The CO 1-0 data have a RMS noise level of 0.8~K at that resolution in our region. For the $^{13}$CO 1-0 data we measured an RMS noise level of 0.4~K.

The \CII{} data were retrieved from the GOTC+ project \citep{Pineda2013} who performed \textit{Herschel/HIFI} observations at a spatial resolution of 12~arcsec at a single pointing \changed{ next to our strip}. Data were provided at main beam temperature scale and a velocity resolution of 0.8~\kms{}.

\section{Results}
\label{sect_results}

\begin{figure*}
   \centering
   \includegraphics[width=8.1cm,angle=270]{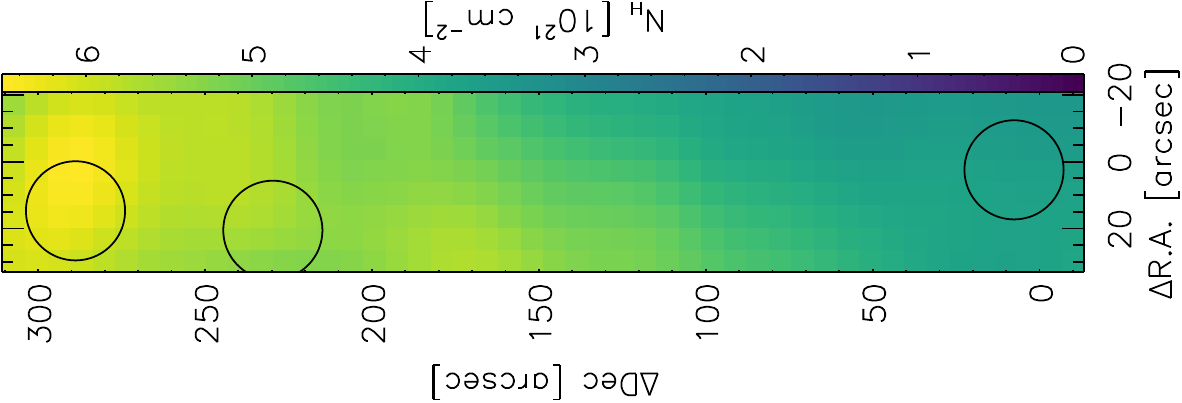}
   \includegraphics[width=8.1cm,angle=270]{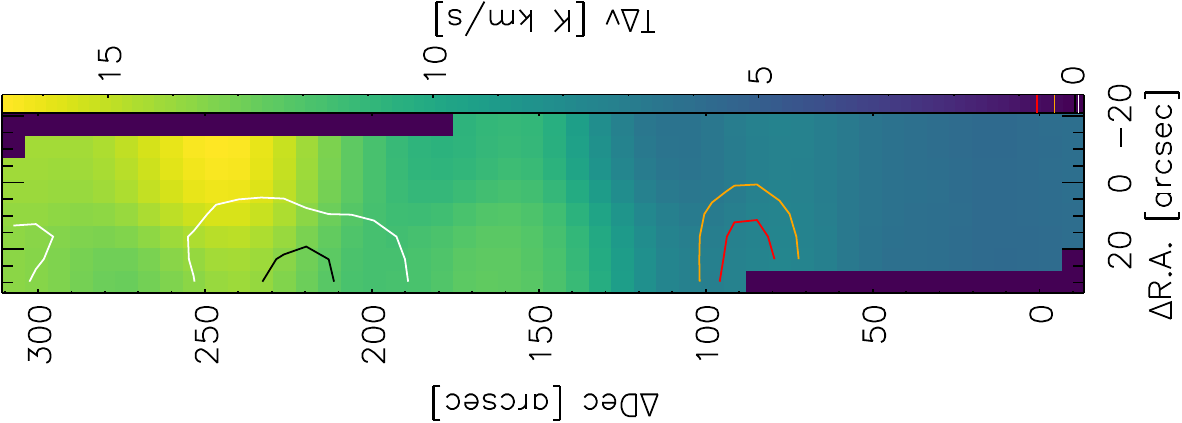}
   \includegraphics[width=8.1cm,angle=270]{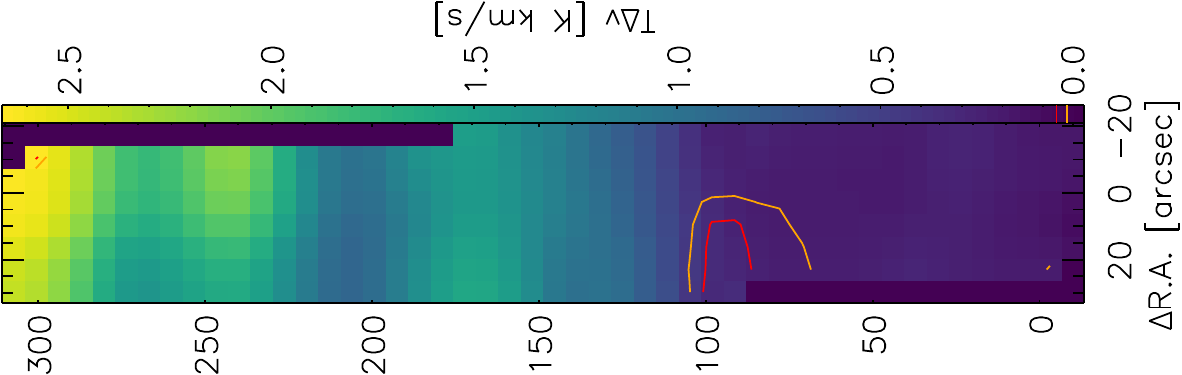}
   \includegraphics[width=8.1cm,angle=270]{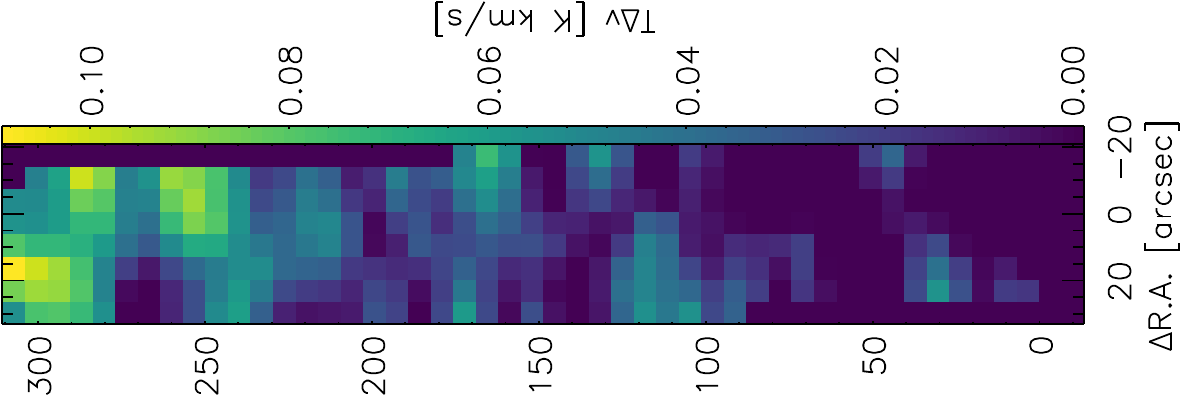}
   \includegraphics[width=8.1cm,angle=270]{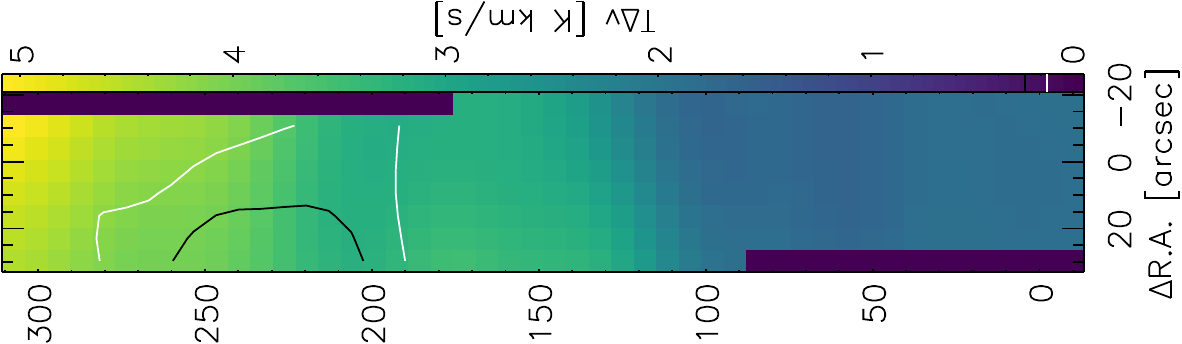}
   \caption{Integrated maps of the observed strip. The left map shows the gas column density derived from the Hi-GAL observations. The circles there indicate the position and \changed{effective} beam size of the spectra discussed in Fig.~\ref{fig_profiles}. The following maps give the line-integrated intensities of CO 2-1, $^{13}$CO 2-1, C$^{18}$O 2-1, and \CI{} 1-0 transitions. Colors represent the integrated intensity for the main velocity component from 11 to 17~\kms{}, black and white contours the integral over the component from 18 to 22~\kms{}, and orange and red contours the integral from 34 to 38~\kms{}. The zero position in the plots is at R.A.(J2000)=7:10:39.8, Dec(J2000)=-11:27:09.}
              \label{fig_maps}%
\end{figure*}

\begin{figure*}
   \centering
   \includegraphics[angle=270,width=0.23\textwidth]{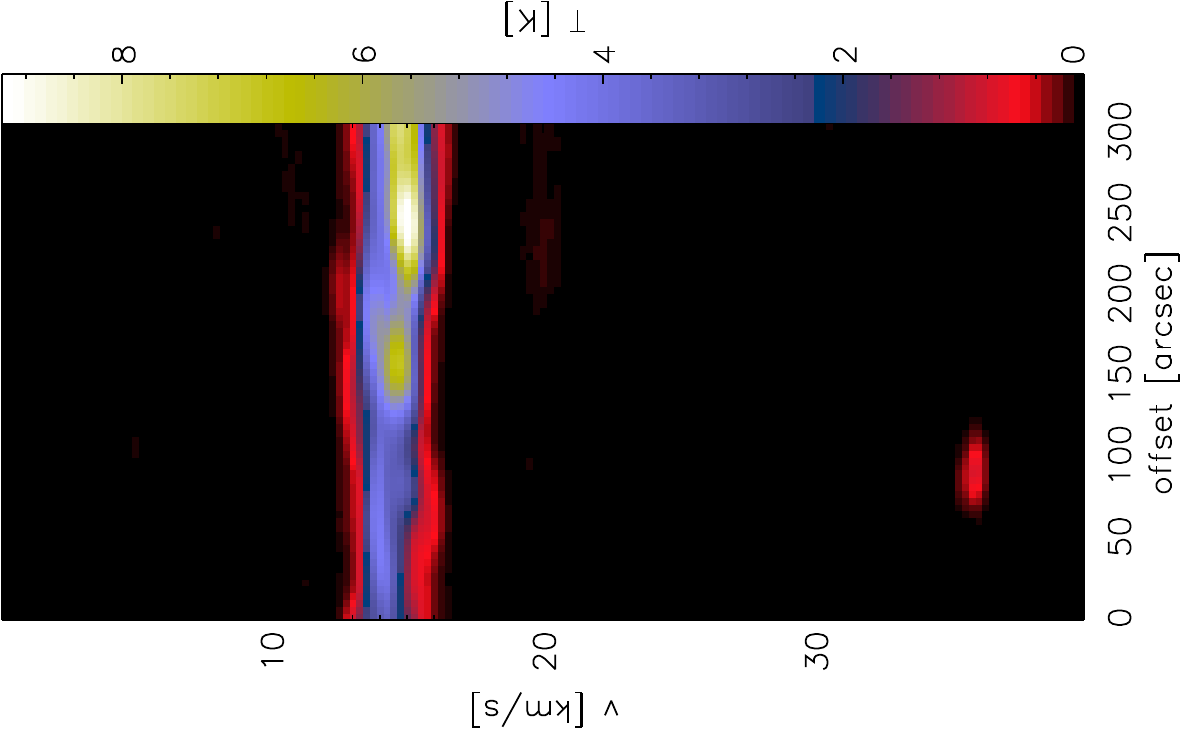}
   \includegraphics[angle=270,width=0.24\textwidth]{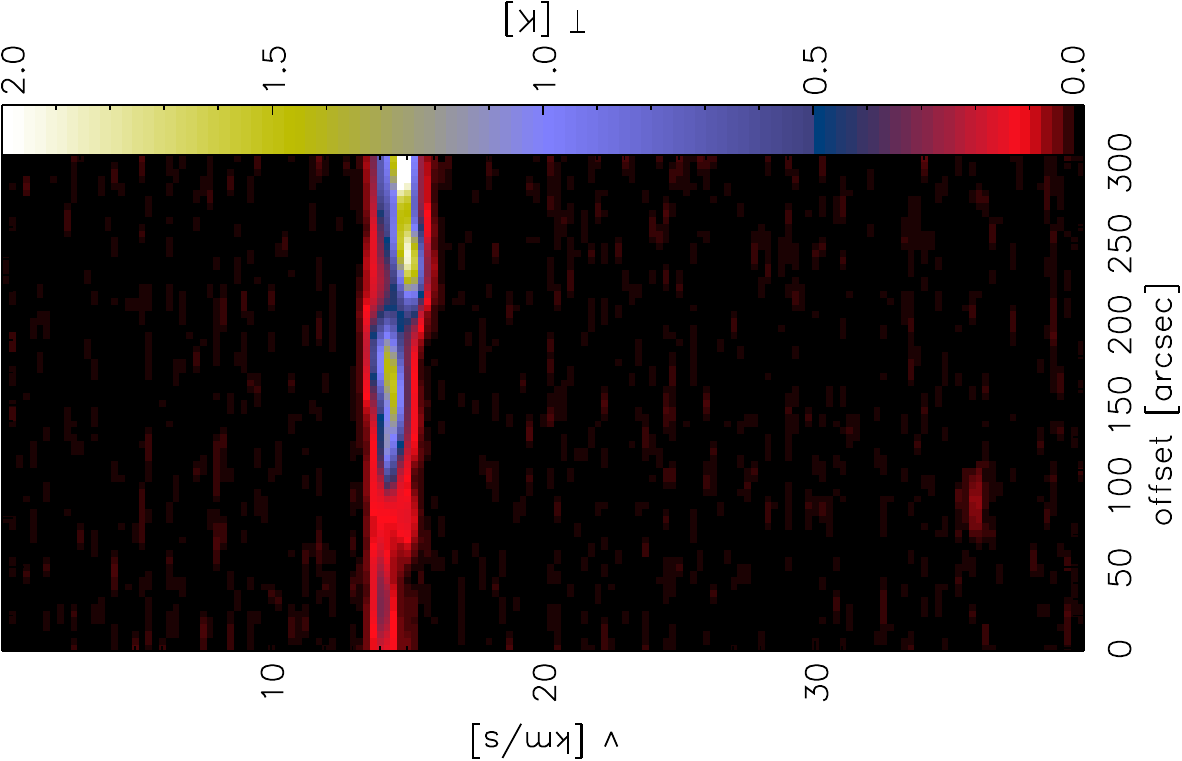}
   \includegraphics[angle=270,width=0.25\textwidth]{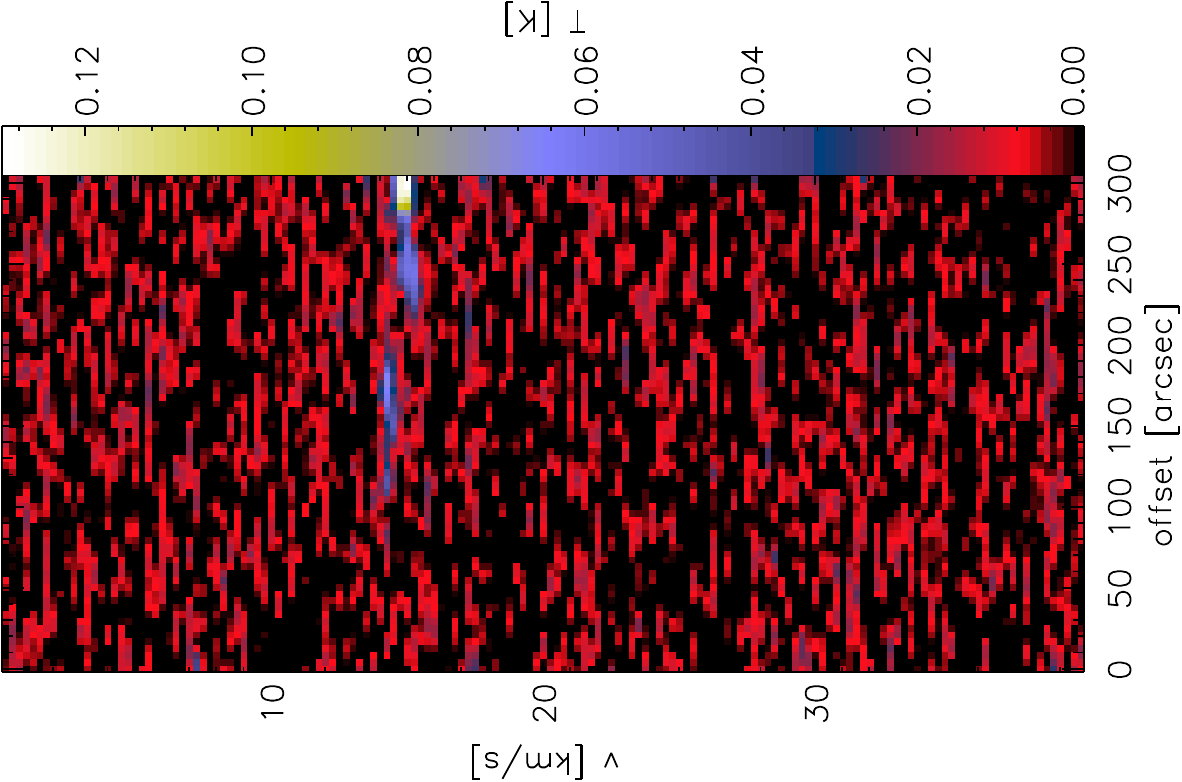}
   \includegraphics[angle=270,width=0.24\textwidth]{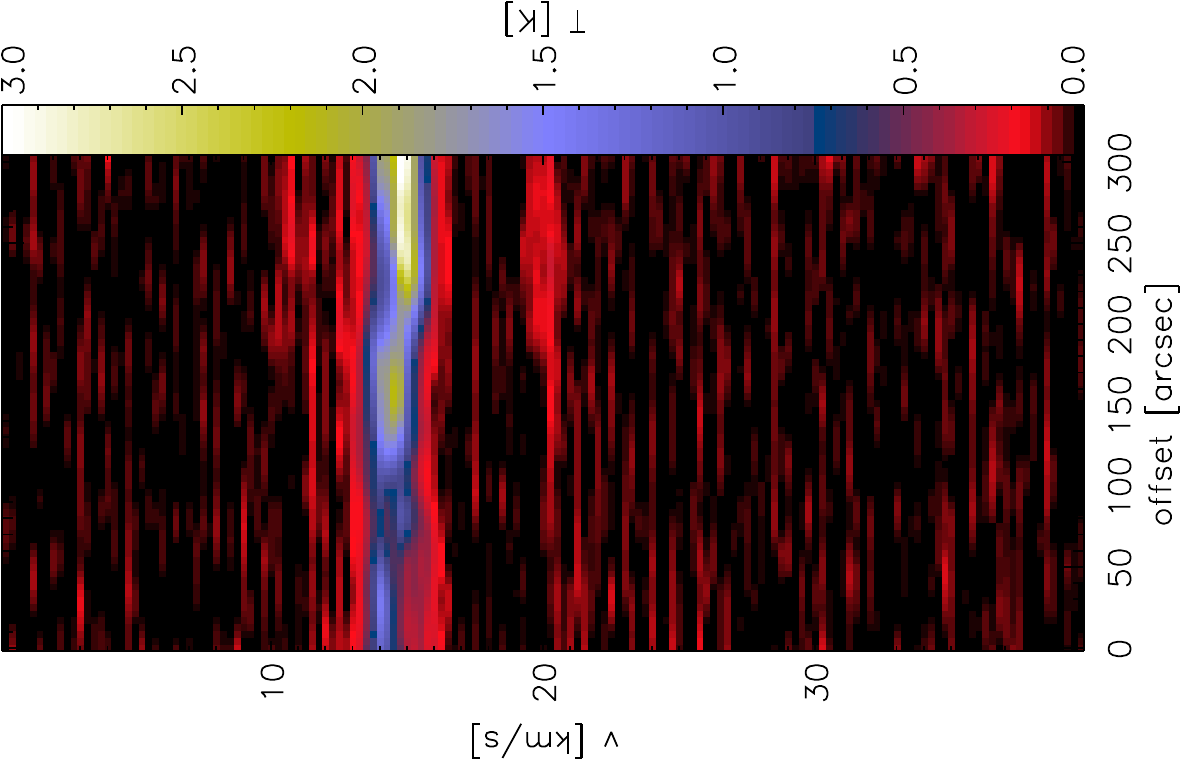}
   \caption{Position-velocity diagrams for the four lines observed: CO 2-1, $^{13}$CO 2-1, C$^{18}$O 2-1, and \CI{} 1-0 (from left to right). All spectra are averaged in the direction perpendicular to the observed strip.
               }
              \label{fig_pvcuts}%
\end{figure*}

Figure~\ref{fig_maps} shows the line integrated intensity maps for the four newly observed transitions, CO 2-1, $^{13}$CO 2-1, C$^{18}$O 2-1, and \CI{} 1-0, in comparison to the dust-based column density map. The different velocity components identified in Fig.~\ref{fig_pvcuts} are shown through different plotting styles. The integrated intensity of the main velocity component is given in colors, black and white contours show the 20~\kms{} component and the orange and red contours show the 36~\kms{} component. Contours are drawn at 50\,\% and 80\,\% of the peak intensity. Lacking contours indicates that the component has not been detected above the noise limit. 

The dynamic range covered by the different tracers is very different. While the dust column density only changes by about a factor of two along the strip, C$^{18}$O 2-1 goes from non-detection at the 10~mK~\kms{} level on one end to 0.45~\Kkms{} in the brightest spots. CO 2-1 and \CI{} 1-0 cover a dynamic range of a factor of about three while $^{13}$CO 2-1 changes by more than a factor ten across the map. Thus, \CI{} 1-0 shows the smoothest spatial distribution among all considered tracers and their transitions. This is confirmed by the high-resolution analysis in Appx.~\ref{appx_highres}. The dust-based column density and the $^{13}$CO 2-1 and C$^{18}$O 2-1 intensity distributions are significantly more patchy. Whether there is a significant difference in that aspect between the $^{13}$CO 2-1 and C$^{18}$O 2-1 maps cannot be judged due to the limited signal to noise in the C$^{18}$O 2-1 data.

%

Fig.~\ref{fig_pvcuts} shows the position-velocity diagrams for the different lines averaged over the pixels perpendicular to the strip. We see the main velocity component around 14~\kms{} in all tracers; however, in C$^{18}$O 2-1 it is only detected at the dense end of the strip. Two other velocity components show up. CO 2-1 and \CI{} show a line at about 20~\kms{}. This is actually more prominent in \CI{} than in CO 2-1. CO 2-1 and $^{13}$CO 2-1 show another component at 36~\kms{} at offsets between 60 and 120~arcsec.

Based on the velocity of the components, we can attribute them to clouds at different kinematic distances using the Galactic rotation curve. From the kinematic distance tool by \citet{Wenger2018tool}, based on the rotation curve from \citet{Reid2014} and the methods described in \citet{Wenger2018}, we obtain a distance of 1.0~kpc for the 14~\kms{} component, a distance of 1.5~kpc for the 20~\kms{} component and a distance of 2.8~kpc for the 36~\kms{} component. At a Galactic longitude of 225~degrees those values need to be multiplied by 0.71 when adding to the solar Galactocentric radius of 8.34~kpc \citep{Reid2014} to obtain the Galactocentric radius of the sources. 

An increase of the Galactocentric radius by up to 2.0~kpc could lead to measurable changes in the metallicity and isotopic ratios in the gas, thereby affecting our analysis of the observations and the expected structure of the PDRs. Using the range of metallicity gradients fitted by  \citet{Cheng2012} and \citet{Lemasle2018}, we find that the metallicity in the Galactic midplane may drop by 17--26\,\% for the 36~\kms{}-component, but only by 6--10\,\% for the 14~\kms{}-component. The drop in metallicity should be reflected in a lower dust content of the gas, leading to a slightly lower dust extinction per gas column, thereby increasing the ratio of the surface tracers, such as \CII{} and \CI{}, relative to the molecular species, such as the CO isotopologues, in the PDRs \citep{Bolatto1999,Roellig2006}. Next to the metallicity, the $^{12}$C/$^{13}$C isotopic ratio changes. From the range of fitted gradients obtained by \citet{Yan2023} and \citet{Milam2005} we see that the $^{12}$C/$^{13}$C ratio may increase by 15--20\,\%
for the 36~\kms{}-component and by 5--7\,\% for the 14~\kms{}-component.
As a consequence, the $^{13}$CO intensity relative to that of CO can be somewhat lower than predicted by PDR models for solar metallicity, both due to the lower self-shielding of $^{13}$CO and the somewhat lower $^{13}$C abundance in fully molecular gas, but the effect should be moderate only.
For the oxygen isotopologues, \citet{Zhang2020} found no significant systematic change on the scales considered here, which is consistent with the theoretical estimates of \citet{Henkel1993}.
Altogether we should consider weak but measurable effects for the 36~\kms{}-component. For most of the emission in the observed strip, which is attributable to the 14~\kms{}-component, the changes should be of the order of the calibration accuracy. Thus, a comparison with models using the ISM abundances at the solar Galactocentric circle is still justified. The same applies to the weak 20~\kms{} component.

\begin{figure*}
   \centering
   \includegraphics[angle=0,width=0.32\textwidth]{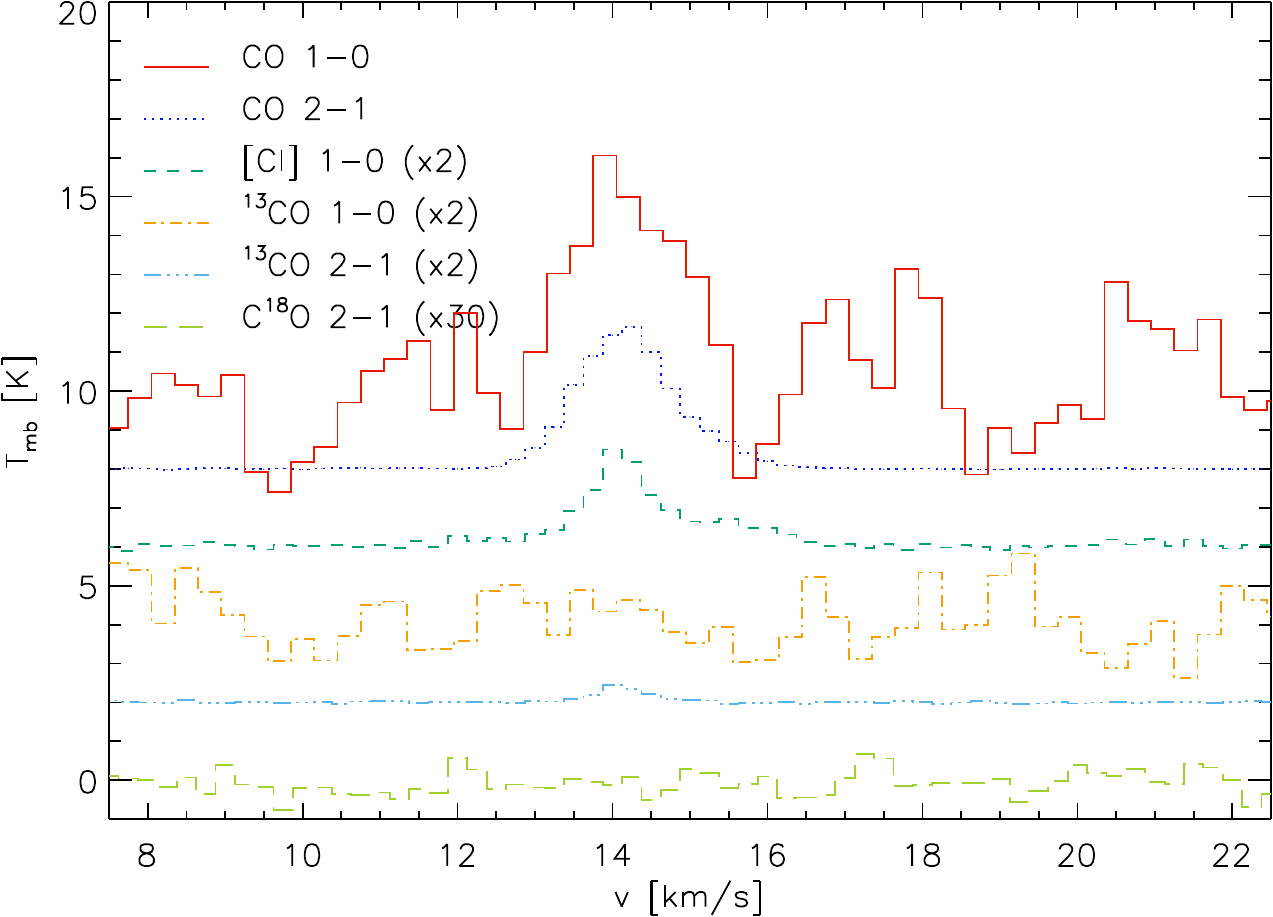}
   \includegraphics[angle=0,width=0.32\textwidth]{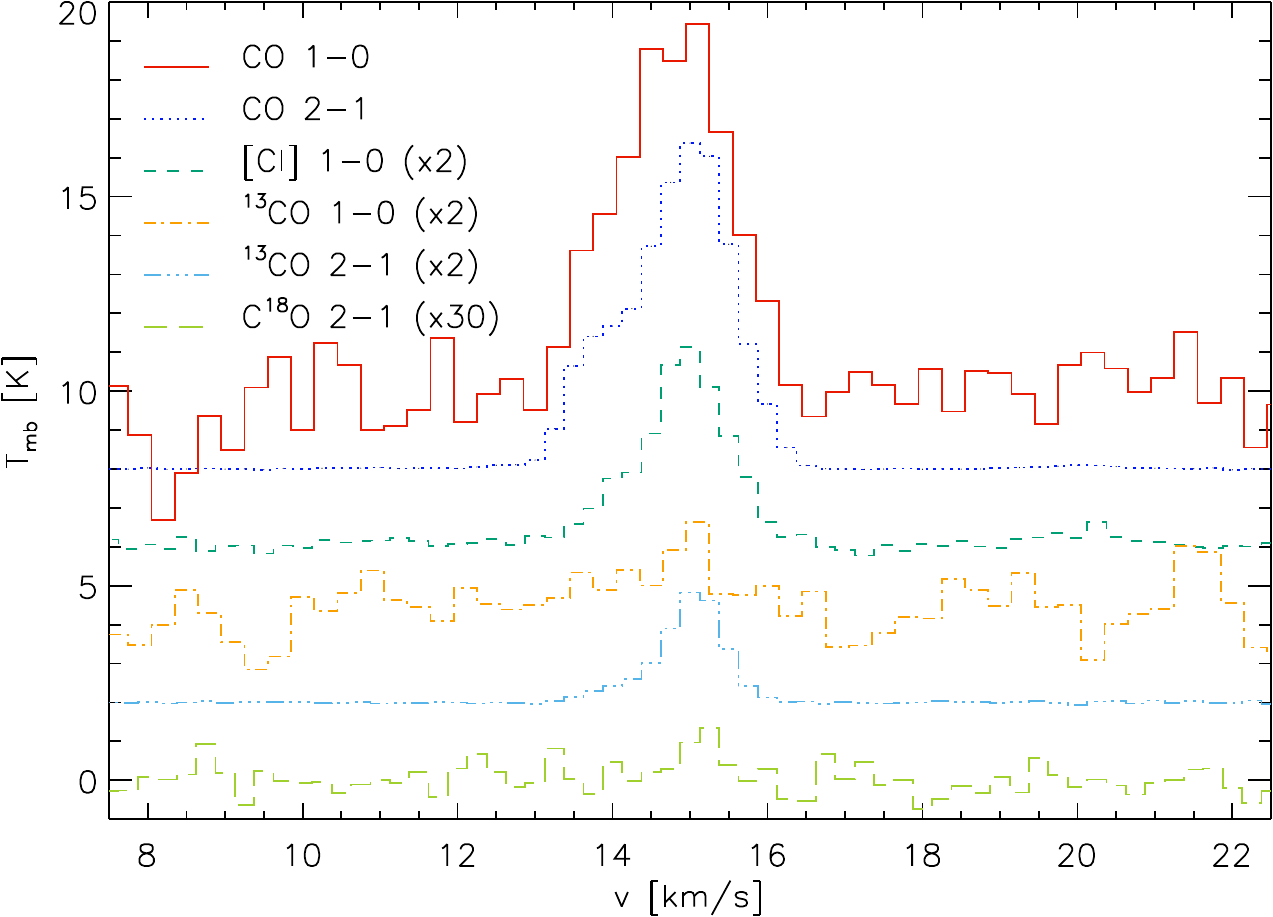}
   \includegraphics[angle=0,width=0.32\textwidth]{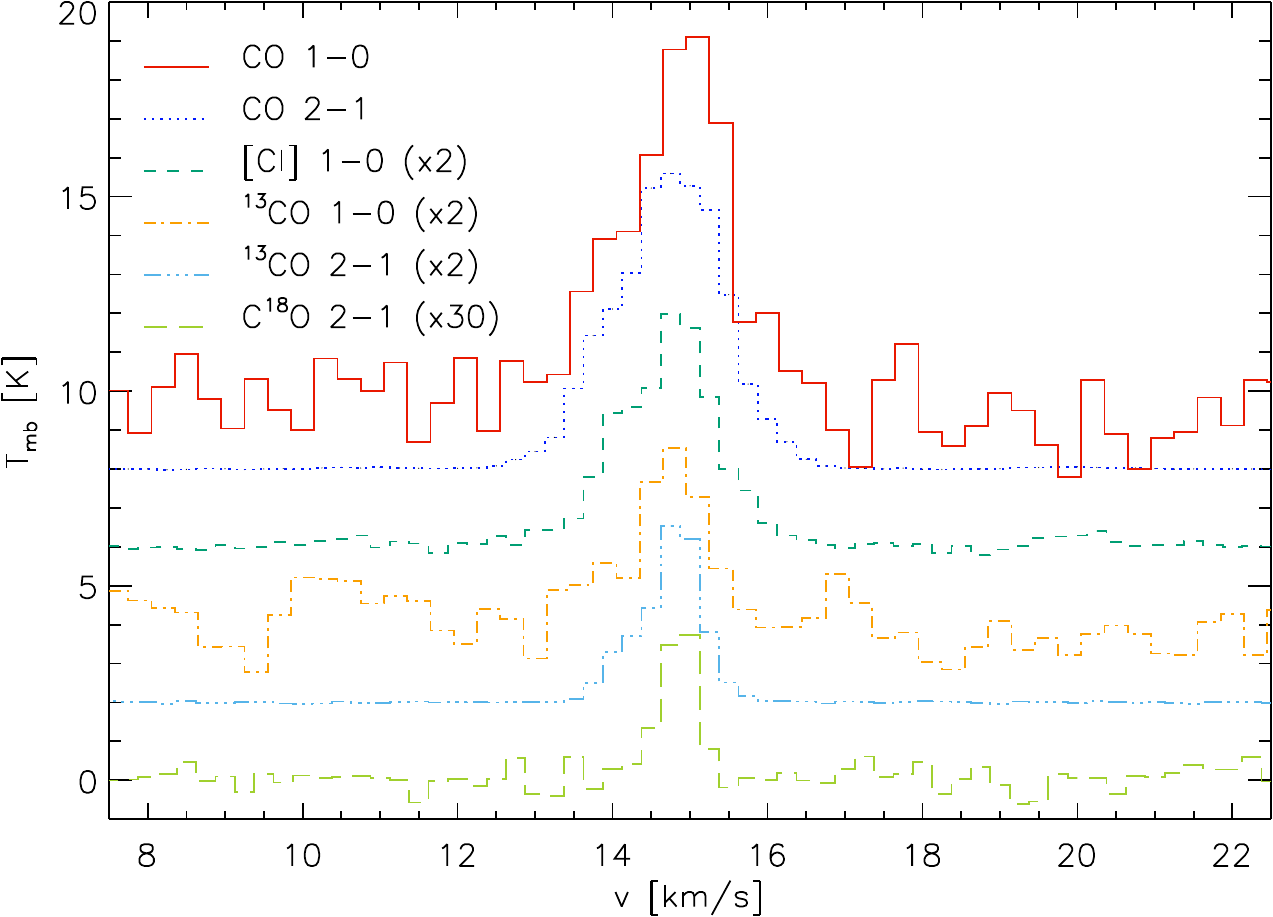}
   \caption{Spectra measured at three selected positions. The left plot is at the zero-position of the map, the central plot at the peak of the 20~\kms{}-component, and the right plot is for the central pixel at the bright end of the strip (see Fig.~\ref{fig_integrated}). The individual spectra have been shifted vertically by 2~K relative to each other to improve the visibility. As the 36~\kms{} component is not detected above the noise level for these three positions the plotted spectral range is limited to the other two velocity components. 
    }
    \label{fig_profiles}%
\end{figure*}

For a comparison of the velocity profiles, we show in Fig.~\ref{fig_profiles} the observed spectra at three selected positions. The left plot shows the spectra at the start of the strip. The central plot gives the spectra at the location of the peak of the 20~\kms{}-component (see Fig.~\ref{fig_maps}), the right plot those for the central pixel at the high column density end of the strip. The plots also show the CO 1-0 and $^{13}$CO 1-0 spectra from the FQS \citep{Benedettini2020}. However, they were measured at a lower resolution of 55~arcsec so that the beam filling can be different from the other lines.

In all three positions, we see a relatively fixed intensity ratio of about two between CO 2-1 and \CI{} 1-0. The line profiles are very similar but we find a hierarchy of line widths. The CO lines have a  typical line FWHM of 1.6~\kms{}, the \CI{} 1-0 line has an FWHM of 1.4~\kms{}l and the $^{13}$CO 2-1 line is narrower, with an FWHM of 0.95~\kms{}, where detected. The C$^{18}$O 2-1 line seems to be even narrower, but this is difficult to quantify due to the low signal-to-noise ratio of the line. It is obvious that the C$^{18}$O 2-1 to $^{13}$CO 2-1 line ratio varies strongly throughout the strip. The $^{13}$CO 1-0 line is somewhat weaker than the 2-1 transition, but it is here again the low signal-to-noise ratio that makes a quantitative comparison difficult. 
The 20~\kms{} component is only visible in CO 2-1 and \CI{} 1-0. \CI{} 1-0 shows it at all three positions, CO 2-1 only at the high column density end of the strip.

\begin{figure}
   \centering \includegraphics[angle=0,width=0.32\textwidth]{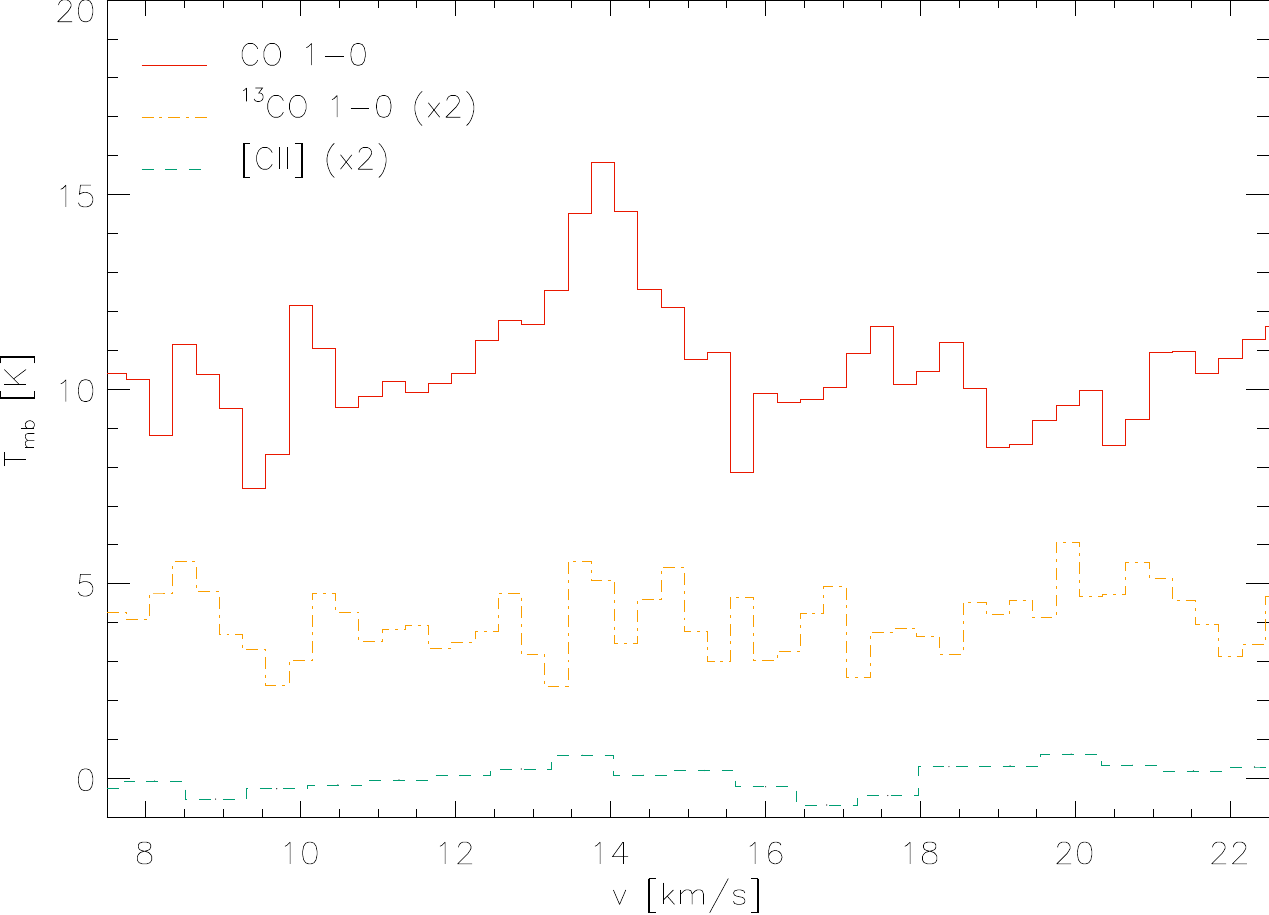}
   \caption{\changed{Spectra of CO 1-0,  $^{13}$CO 1-0, and \CII{} at the position of the GOTC+ observations next to our strip. The CO and $^{13}$CO 1-0 profiles have been shifted in the same way as in Fig.~\ref{fig_profiles} to allow for a direct comparison.} 
    }
    \label{fig_profiles_cii}%
\end{figure}   

\changed{To judge to what degree the \CII{} GOTC+ spectrum \citep{Pineda2013} can be taken as representative for our strip, we show in Fig.~\ref{fig_profiles_cii} the spectra of CO 1-0 and $^{13}$CO 1-0 for the GOTC+ position together with the \CII{} spectrum. One has to keep in mind that they are} measured at a different resolution so that \changed{they cannot be quantitatively compared.} 
In \CII{}, we only have a non-detection or tentative detection with a maximum line integrated intensity of 0.2~\Kkms{}. This corresponds to 1.4 times the baseline noise at the resolution of the line width of 1.5~\kms{}. \changed{The profile of CO 1-0 with an integrated intensity of 8~\Kkms{}, the non-detection of $^{13}$CO 1-0, and the dust-based column density of $3.4 \times 10^{21}$\pccm{} all match the observations within the lower 80~arcsec of our strip. Hence, we consider the upper limit  of 0.2~\Kkms{} for the \CII{} intensity as representative for the low-column-density end of our strip.}

\section{Correlations}
\label{sect_ratios}

\begin{figure*}
   \centering
   \setlength{\unitlength}{1mm}
   
   \hspace*{2mm}
   \begin{overpic}[height=4.7cm]{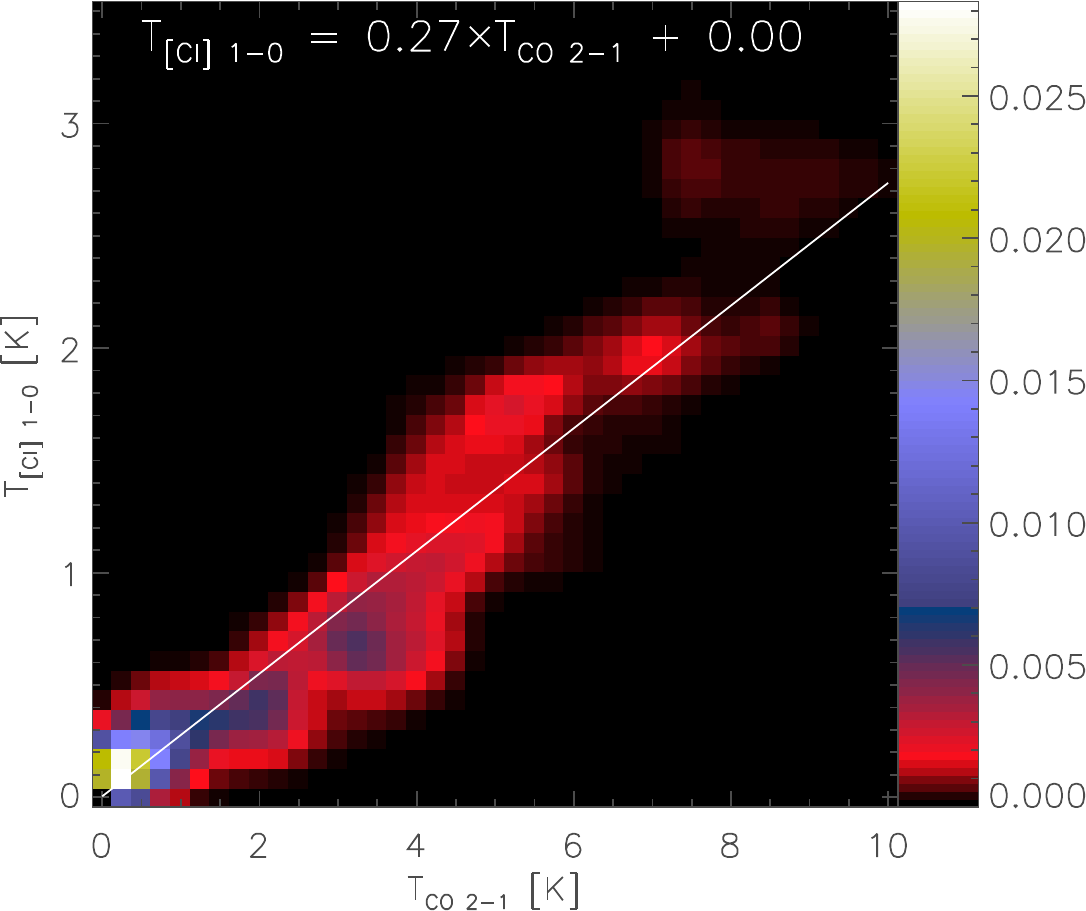}
      \put(-3,2){\textbf{(a)}}
   \end{overpic}\hspace*{3mm}
   \begin{overpic}[height=4.7cm]{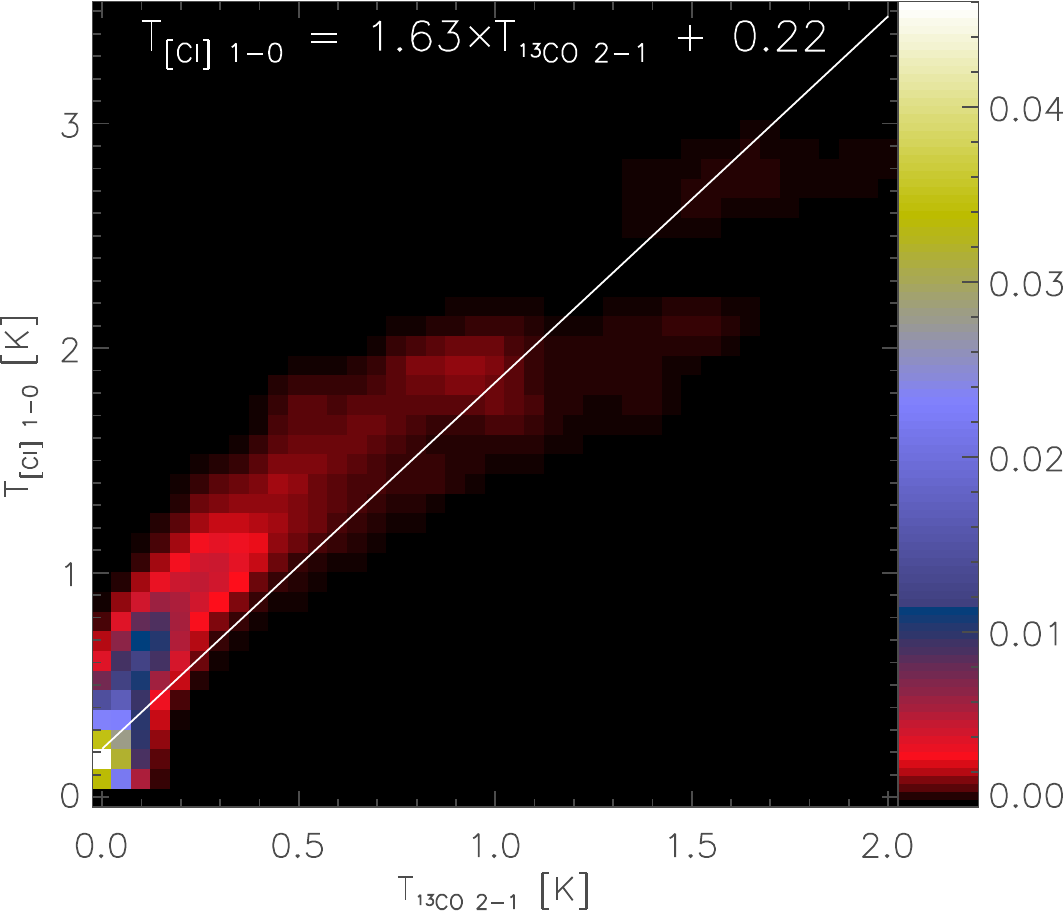}
      \put(-3,2){\textbf{(b)}}
   \end{overpic}\hspace*{3mm}
   \begin{overpic}[height=4.7cm]{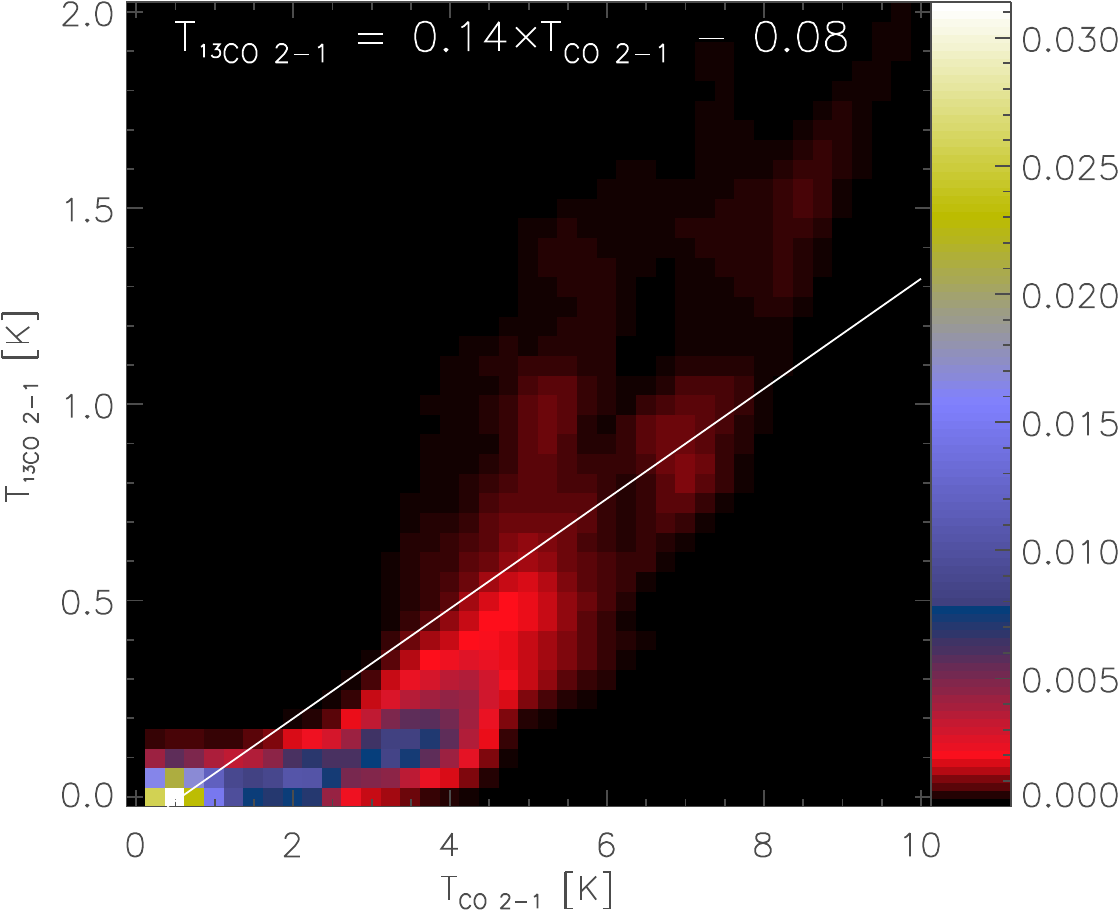}
      \put(-1,2){\textbf{(c)}}
   \end{overpic}\vspace*{2mm}\\
   
   \begin{overpic}[height=4.7cm]{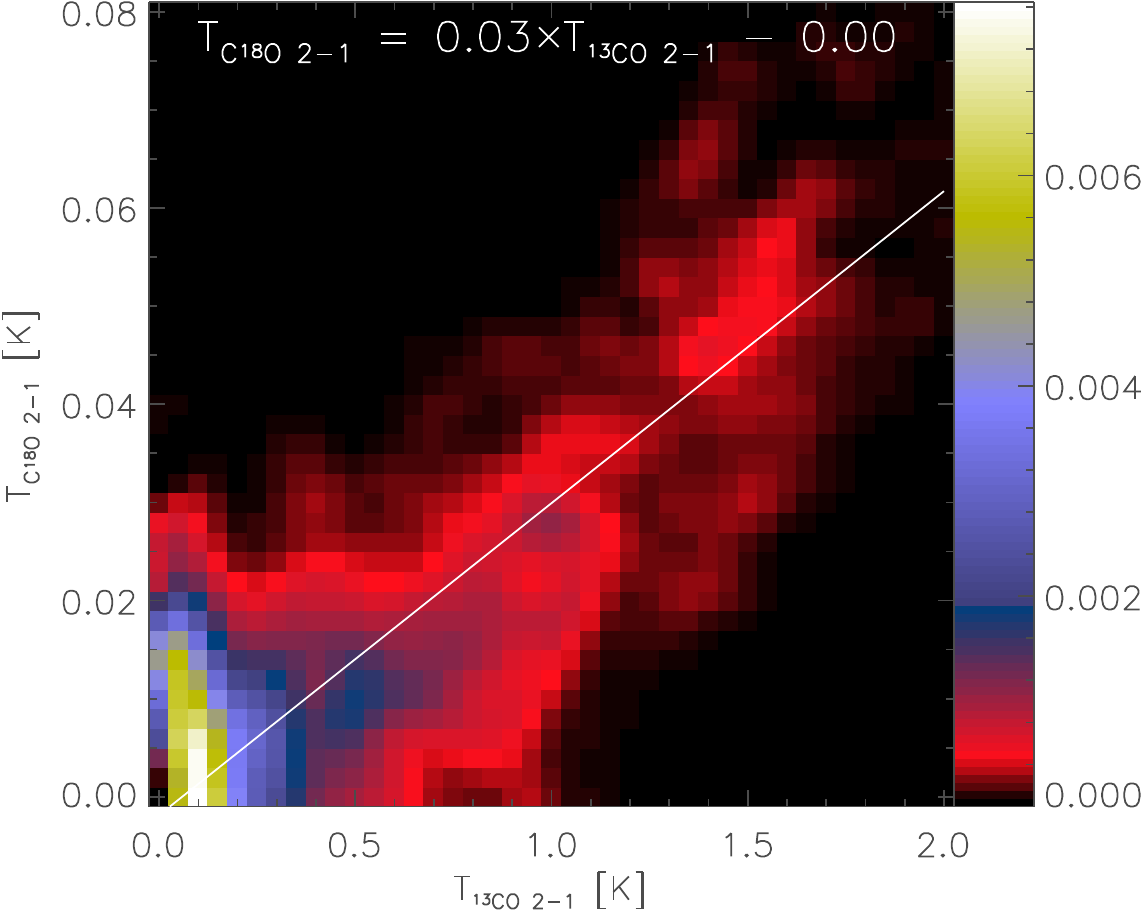}
      \put(3,2){\textbf{(d)}}
   \end{overpic}
   \begin{overpic}[height=4.7cm]{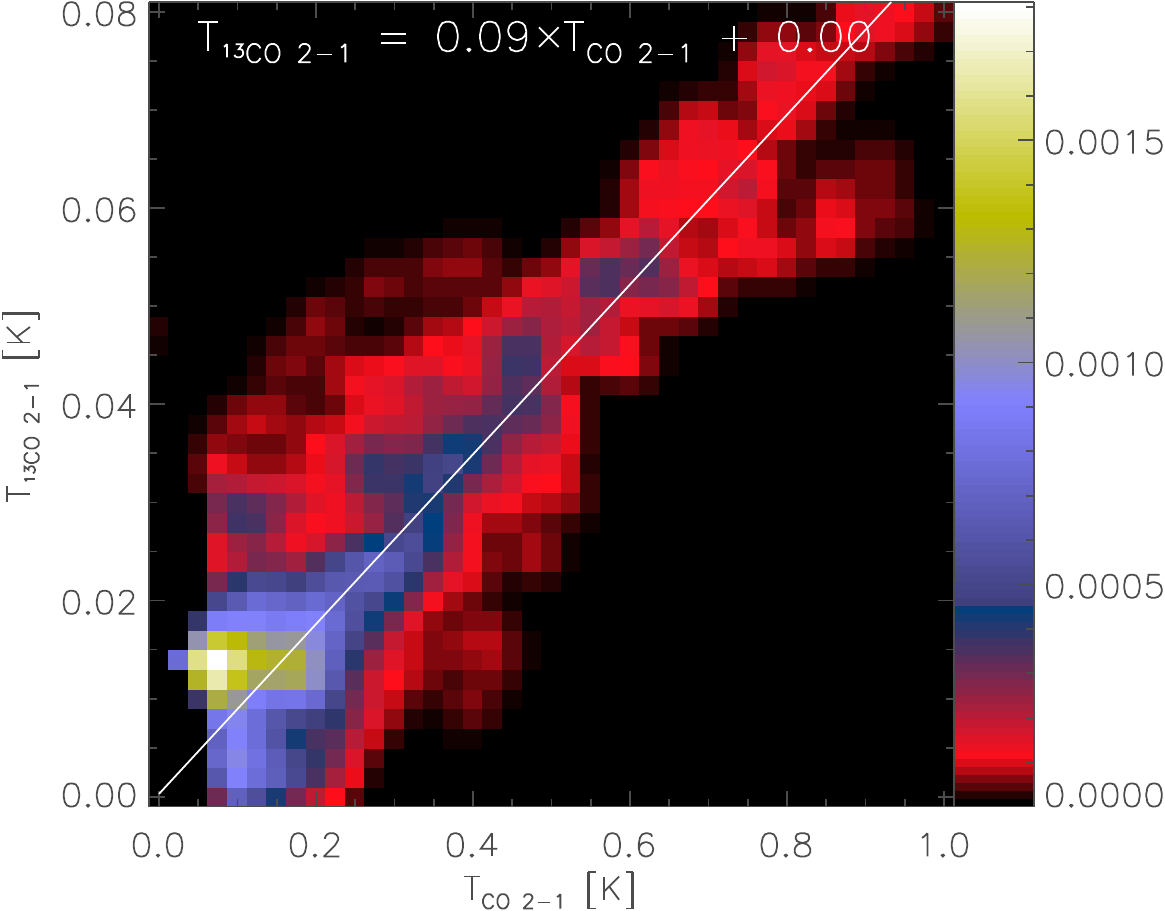}
      \put(3,2){\textbf{(e)}}
   \end{overpic}
   \begin{overpic}[height=4.7cm]{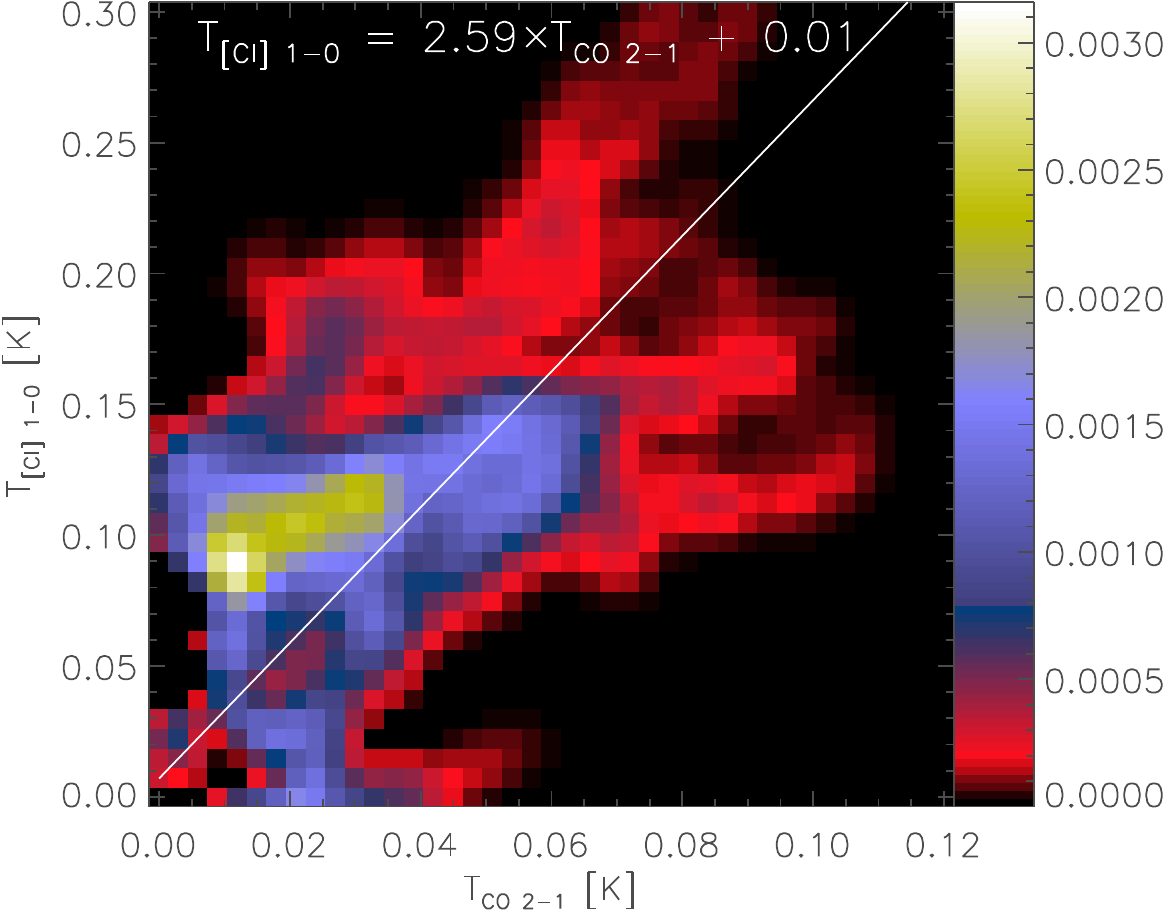}
      \put(1,2){\textbf{(f)}}
   \end{overpic}
   \caption{Distribution of line intensities comparing different transitions at the same spatial-spectral pixel. The panels compare  \textbf{a:} CO 2-1 to [CI] 1-0, \textbf{b:} $^{13}$CO 2-1 to [CI] 1-0, \textbf{c:} CO 2-1 to $^{13}$CO 2-1, \textbf{d:}  $^{13}$CO 2-1 to C$^{18}$O 2-1, for the 14~\kms{} component, i.e. all spectral pixels in the line window from 11 to 17~\kms{}, and \textbf{e:} 
   CO 2-1 to $^{13}$CO 2-1 in the 36~\kms{} component (34-38~\kms{}), and \textbf{f:} CO 2-1 to [CI] 1-0 for the 20~\kms{} component (18-22~\kms{}). The noise distribution determined outside of the spectral window is subtracted to emphasize the relation for true emission only.
   The white line and the equation on top show the results of a linear fit to all noise-weighted points forming the distributions.}
    \label{fig_corr_main}
\end{figure*}

To verify whether our \CI{} data follow the often observed tight correlation to the $^{13}$CO emission, we studied the correlation between the \CI{} emission and the other tracers within the position-velocity cubes. Figure~\ref{fig_corr_main} shows the distribution of intensities within the frequency windows of the velocity components with significant emission. The first four panels cover the main emission component between 11 and 17~\kms{}. The last two panels show the corresponding distributions for the two other velocity components in the lines that were detected above the noise level. The distributions are shown as probability density functions (PDFs) of the spectral pixel intensities where we have subtracted the PDFs of the noise contribution by measuring the latter in channels outside of the line window. Therefore, the plots should represent actual emission only.

Looking at the main velocity component, it becomes obvious that the relation between $^{13}$CO 2-1 and \CI{} 1-0 (second panel) is not at all linear while the relation between CO 2-1 and \CI{} 1-0 (first panel) follows a linear trend in spite of very different optical depths (see Sect.~\ref{sect_lte}). This is consistent with the relation between $^{13}$CO 2-1 and CO 2-1 being non-linear. The relation between $^{13}$CO 2-1 and C$^{18}$O 2-1 also follows a linear trend, but there seems to be some excess of C$^{18}$O 2-1 at a level of up to $2\sigma\sub{noise}$ for low levels of $^{13}$CO 2-1. In contrast to the 14~\kms{} component, the 36~\kms{} component shows a quite linear trend between $^{13}$CO 2-1 and CO 2-1. For the 20~\kms{} component, it is remarkable that we find significant \CI{} 1-0 emission at a $3.5\sigma\sub{noise}$ level in spectral pixels with no or very little CO 2-1 emission, indicating the presence of CO-dark gas with atomic carbon.

To characterize typical line ratios, we fitted the relations between the different intensities by a linear model accounting for uncertainties in both axes using the LINFITEX function \citep{Numrec} implemented in the MPFIT package\footnote{http://purl.com/net/mpfit} from \citet{Markwardt2009}. The results are given in Table~\ref{tab_corr} and the fits are overplotted in Fig.~\ref{fig_corr_main}. It is clear that apart from the CO 2-1/\CI{} 1-0 relation, linear fits are no good representation of the observed behavior. The uncertainty of the linear fit is typically very small due to the large number of spectral pixels compared but it actually gives no indication for the quality of a linear model. Therefore, we do not only give the parameter uncertainties from the fit but also computed the scatter of the measured intensities around the fitted line. This is expressed as the intercept and slope of a linear description of the standard deviation of the distribution of values around the fitting line.

Next to the relations for the line ratios, Table~\ref{tab_corr} also shows the linear fit results when comparing the lines with the total column density. Here, the line intensities are integrated over all velocity components.

\begin{table*}
      \caption[]{Linear fit parameters for the relations between the different line intensities.}
         \label{tab_corr}
         \begin{tabular}{p{2.5cm}p{1.7cm}p{1.5cm}rrrrrr}
            \hline
            Independent & dependent & component & \multicolumn{3}{c}{slope} &  \multicolumn{3}{c}{intercept} \\
            intensity & intensity && value & uncertainty & scatter & value & uncertainty & scatter\\
            \hline
CO 2-1 & \CI{} 1-0 & 14~\kms{} & 0.27 & 0.00 & 0.04 & 0.004 & 0.001 & 0.058 \\
CO 2-1 & \CI{} 1-0 & 20~\kms{} & 2.59 & 0.03 & -0.20 & 
0.007 & 0.001 & 0.059 \\
$^{13}$CO 2-1 & \CI{} 1-0 & 14~\kms{} & 1.63 & 0.00 & 0.11 & 0.216 & 0.001 & 0.129 \\ 
CO 2-1 & $^{13}$CO 2-1 & 14~\kms{} &  0.14 & 0.00 & 0.05 &  -0.081 & 0.000 & -0.002 \\
CO 2-1 & $^{13}$CO 2-1 & 36~\kms{} & 0.086 & 0.004 & -0.008 &   0.0003 & 0.0002 & 0.0106 \\
CO 2-1 & C$^{18}$O 2-1 & 14~\kms{} & 0.0038 & 0.0001 & 0.0011 &    -0.0032 & 0.0001& 0.0105 \\
$^{13}$CO 2-1 & C$^{18}$O 2-1 & 14~\kms{} &
0.032 &  0.000 &  0.001 & -0.0019 & 0.0001& 0.0111\\
            \hline
$N\sub{H}$ [$10^{21}$~cm$^{-2}$] & \multicolumn{2}{l}{CO 2-1 [K~\kms{}]}   &   3.98 & 0.11 & 0.14 &  -9.11 &  0.55 & -0.01 \\
$N\sub{H}$ [$10^{21}$~cm$^{-2}$] & \multicolumn{2}{l}{$^{13}$CO 2-1 [K~\kms{}]}   &  0.80 &  0.02 & 0.05 &  -2.90 &  0.11 &   -0.09\\
$N\sub{H}$ [$10^{21}$~cm$^{-2}$] & \multicolumn{2}{l}{C$^{18}$O 2-1 [K~\kms{}]}   & 0.034 & 0.001& 0.003 & -0.144& 0.006 &  0.001\\
$N\sub{H}$ [$10^{21}$~cm$^{-2}$] & \multicolumn{2}{l}{\CI{} 1-0 [K~\kms{}]}   & 1.12& 0.03& 0.03 & -2.60 & 0.15 & -0.03\\
\hline
     \end{tabular}
     \tablefoot{The mutual comparison between line intensities are performed channel by channel. The comparison to the dust column is done for the line integrated intensities.}
\end{table*}

When looking at the line intensity ratios, only two relations have a non-zero intercept within the uncertainties, the relations between $^{13}$CO 2-1 and either CO 2-1 or \CI{} 1-0. This shows that a significant intensity of CO 2-1 and \CI{} 1-0 has to build up, before the formation of $^{13}$CO sets in. A similar effect is expected for C$^{18}$O, but there the uncertainties are larger due to the weak detection.

For the relations of the line-integrated intensities to the dust-based column, we can translate the intercept to the column density for which the line intensity becomes positive by dividing the negative intercept by the slope. In this way, we obtain a minimum column density for the formation of CO and \CI{} of $2.3\times 10^{21}$~cm$^{-2}$, for $^{13}$CO of $3.6 \times 10^{21}$~cm$^{-2}$, and for C$^{18}$O of $4.2 \times 10^{21}$~cm$^{-2}$. Because of this intercept, the slope of the fitted line must not be mixed with the typical ratio between the two quantities that is discussed below. As the line starts only above the formation threshold, the slope is between two and almost ten times higher than the mean ratio over the map (Table~\ref{tab_integrated}).

\section{Modelling}
\label{sect_models}

\subsection{Line ratios}

  \begin{figure}
   \centering
   \includegraphics[width=0.9\columnwidth]{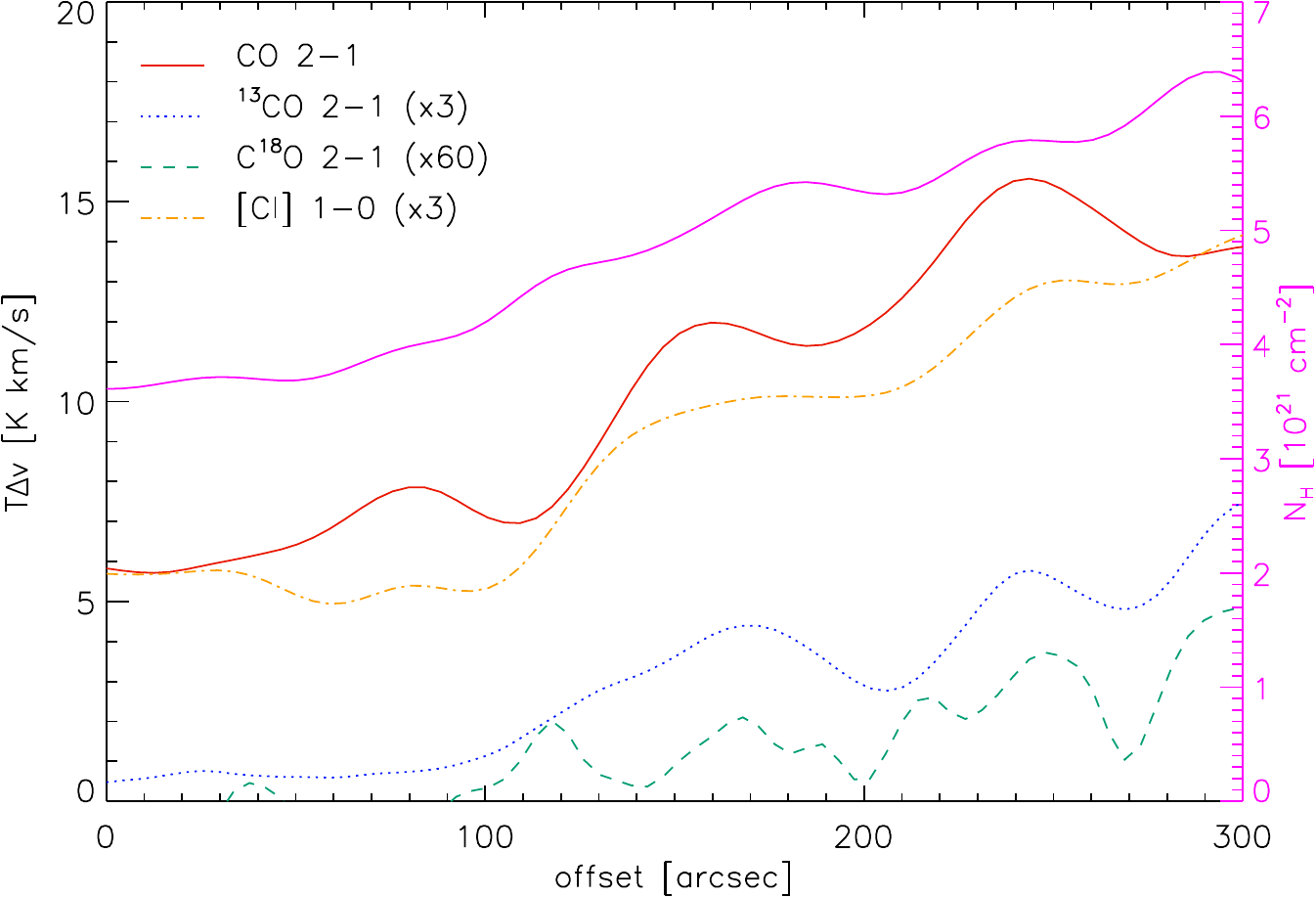}
      \caption{Comparison of the line integrated intensities of \CI{} and the CO isotopologues (left axis) with the column density along the strip (right axis, purple). All data are averaged perpendicular to the strip direction.
     }
         \label{fig_integrated}
   \end{figure}

\begin{table}
      \caption[]{Mean, minimum, and maximum values for column density and the  different intensities and intensity ratios integrated over the main velocity component and averaged perpendicular to the observed strip.} 
         \label{tab_integrated}
     \begin{tabular}{lrrr}
         \hline
         ratio  & mean & minimum & maximum \\
         \hline
$^{13}$CO 2-1 / CO 2-1 & 0.085 & 0.027 & 0.182 \\
C$^{18}$O 2-1 / CO 2-1 & 0.0014 & 0 & 0.0059 \\
\CI{} 1-0 / CO 2-1 & 0.28 & 0.23 & 0.33 \\
\hline
\CII{} / CO 2-1\tablefootmark{a} & & 0 & 0.035 \\
$^{13}$CO 1-0 / CO 2-1 & 0.15 & 0.05 & 0.24 \\
CO 1-0 / CO 2-1 & 1.25 & 1.10 & 1.55 \\
\hline
\multicolumn{2}{l}{ratio [$\Kkms{}/(10^{21}~\mathrm{cm}^{-2})$]}\\
\hline
CO 2-1 / $N\sub{H}$  &  2.1 & 1.5 & 2.7 \\
$^{13}$CO 2-1 / $N\sub{H}$  &
 0.19 & 0.05 & 0.40 \\
C$^{18}$O 2-1 / $N\sub{H}$  &
 0.0034 & 0 &  0.013 \\
\CI{} 1-0 / $N\sub{H}$  &
 0.57 & 0.41 & 0.74\\ 
\hline
\CII{} / $N\sub{H}$ & 
& 0 & \changed{0.059} \\
$^{13}$CO 1-0 / $N\sub{H}$ &  0.32 & 0.09 & 0.54 \\
CO 1-0 / $N\sub{H}$ & 2.6 & 2.1 & 3.2 \\
\hline
values\\
\hline
$N\sub{H}$ [$10^{21}~\mathrm{cm}^{-2}$] & 4.9 & 3.5 & 6.5 \\
CO 2-1 [$\Kkms{}$] & 10.3 & 5.7 & 15.5\\
\hline
\end{tabular}\\
\tablefoot{The last three lines of each ratio are not obtained at the same telescope resolution so that they are only valid if the emission is extended.\\
\tablefoottext{a}{\changed{As discussed in Sect.~\ref{sect_results} we used the \CII{} data for the low-column-density end of our strip.}}}
\end{table}

To determine the physical parameters of the main gas component we consider intensities averaged in the direction perpendicular to the strip and integrated over the 14~\kms{} line. They are plotted in Fig.~\ref{fig_integrated} together with the column density profile from the dust observations. As already discussed in Sect.~\ref{sect_results} the \CI{} line varies least across the strip, being similar to the dust profile, while all molecular lines show stronger variations with bumps in the CO isotopologue lines around 160 and 250~arcsec offset and significant emission in $^{13}$CO 2-1 and C$^{18}$O 2-1 only at offsets above 100~arcsec.

For an estimate of the column of the different species we consider the intensities integrated over the main velocity component. For a physical model of the gas in terms of a PDR,  we do not fit absolute intensities, but consider ratios, characterizing the line emission efficiency of a gas column. One the one hand we use the line intensities relative to the total gas column, $N\sub{H}$. One the other hand we use the intensities relative to the CO 2-1 line intensity the characterize the emission efficiency of the molecular gas only. In Table~\ref{tab_integrated}, we list the mean, minimum and maximum values and ratios along the strip used in the subsequent modelling, as they could be read from Fig.~\ref{fig_integrated}.

\subsection{Uniform slab approximation}
\label{sect_lte}

\begin{figure*}
   \centering
   \includegraphics[angle=0,height=4.2cm]{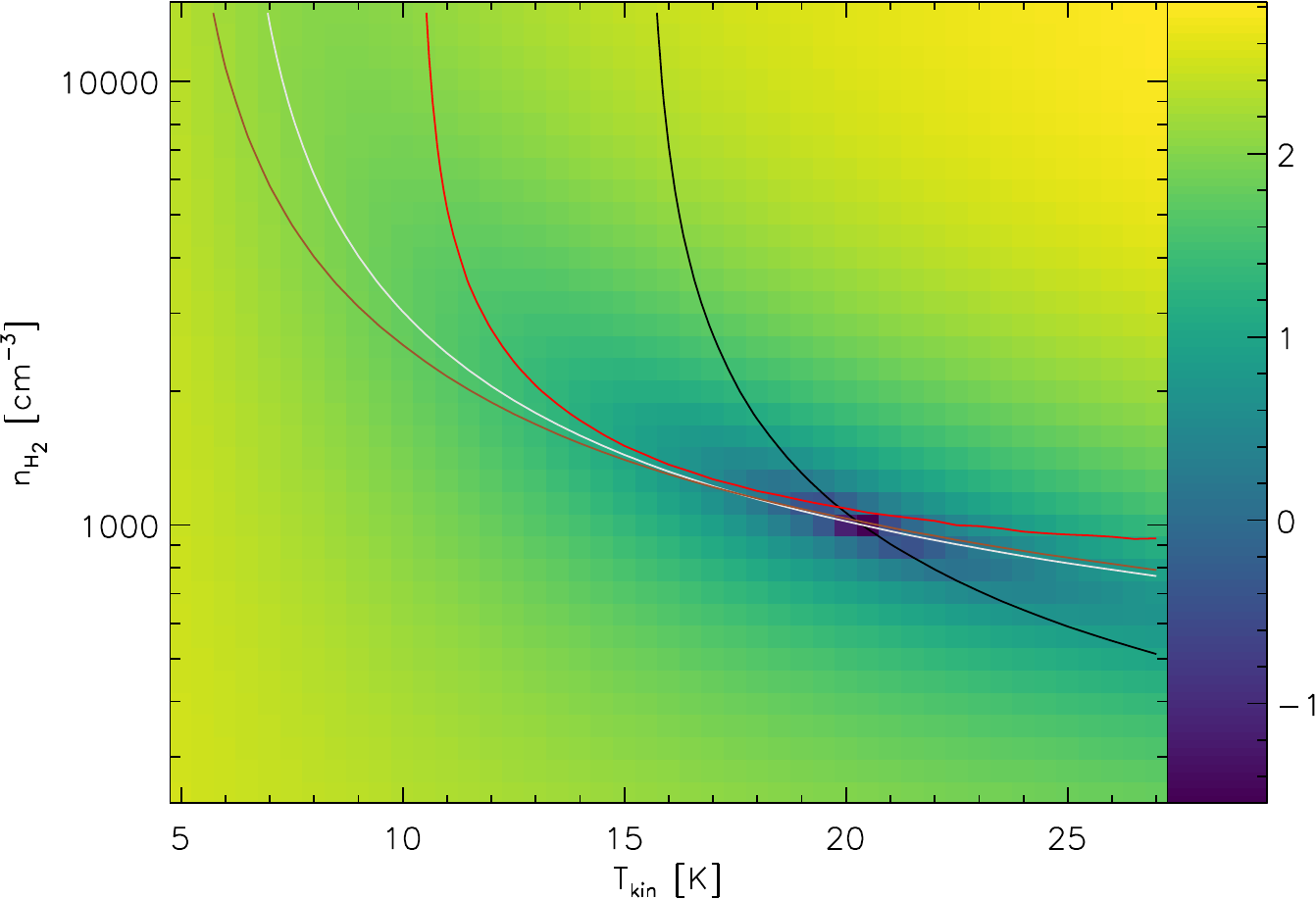}
   \includegraphics[angle=0,height=4.2cm]{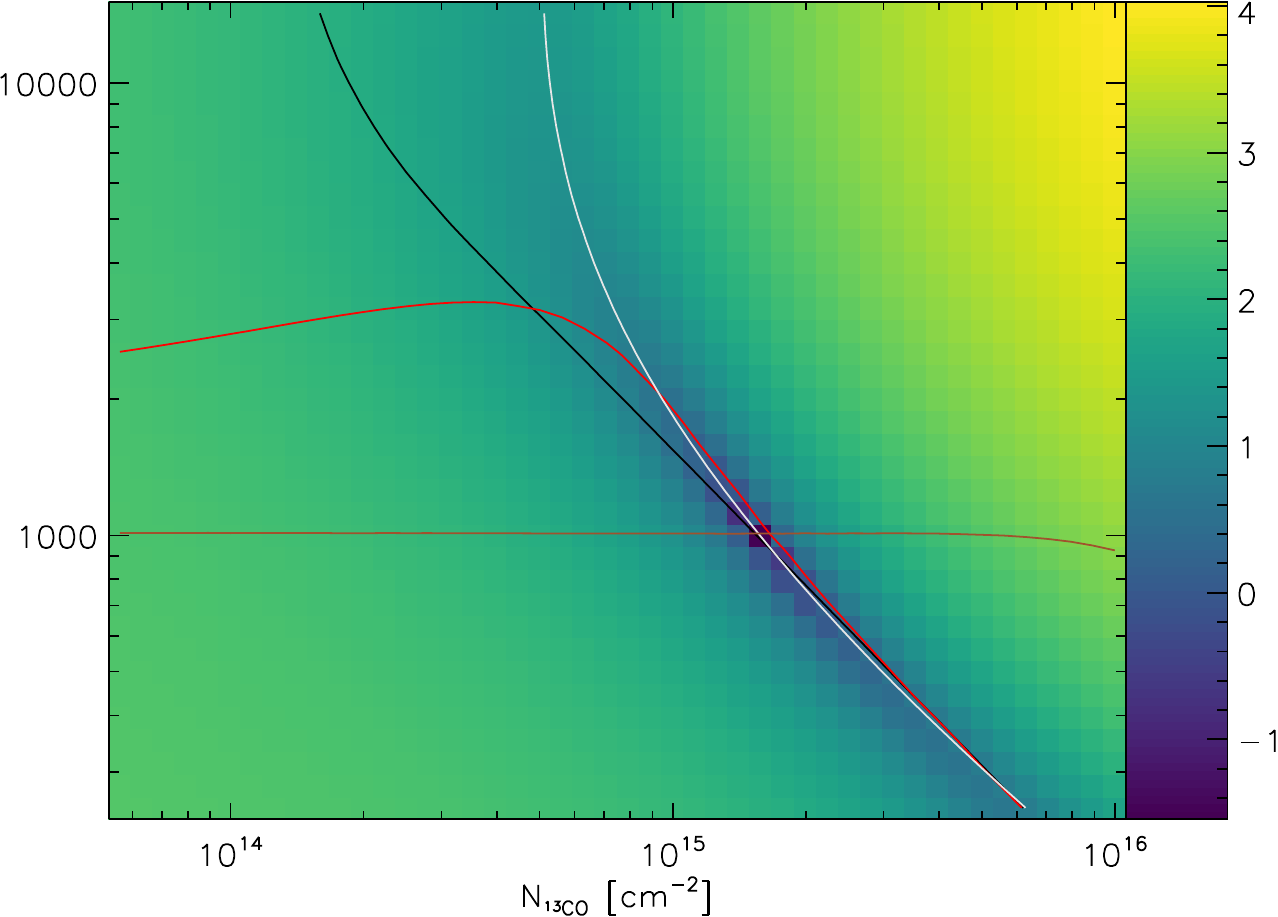}
   \includegraphics[angle=0,height=4.2cm]{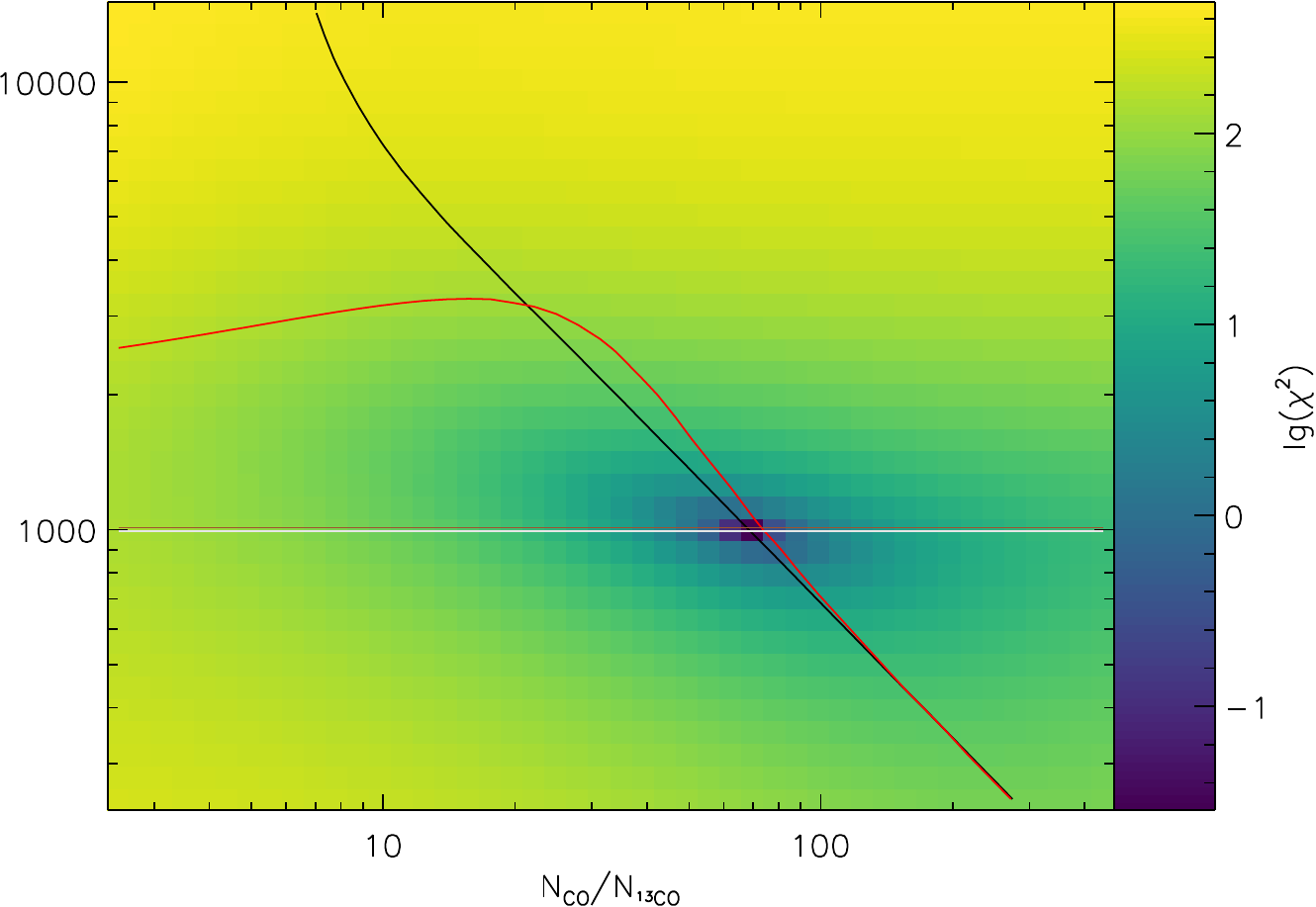}
   \caption{Two-dimensional cuts through the $\chi^2$ topology of the RADEX simulation of the observed intensities of the CO 1-0, 2-1 and $^{13}$CO 1-0 and 2-1 lines. The $\chi^2$ values are given in logarithmic units, so that only the dark-blue area indicates a valid fit. All slices are taken through the $\chi^2$ minimum. The contour lines show the parameters where the observed values were matched. Black and white  lines give the integrated intensities of CO 2-1 and $^{13}$CO 2-1, and red and brown lines show the ratios of CO 1-0/CO 2-1 and $^{13}$CO 1-0/$^{13}$CO 2-1.}
    \label{fig_radexfit}
\end{figure*}

In a first approach we try to analyze the observed intensities in the frame of a uniform medium characterized by the parameters density, kinetic temperature, column density within the telescope beam, beam filling factor, and the abundance of the individual species. The total column density of gas is constrained by the dust observations, but this still leaves three free parameters plus one for each observed molecule. Unfortunately, our observational constraints from the line intensities are insufficient to derive this number of parameters as we have at maximum two lines per species. Hence, we have to make some simplifying assumptions. The general assumption that we apply here is a beam filling of unity, this means no breaking up of the material on the scale below the telescope beam. This is well justified by the smoothness of the intensity distribution in all tracers and the smooth distribution of the \CI{} map at the even higher resolution shown in Appx.~\ref{appx_highres}. 

The next approximation can be made by assuming local thermodynamic equilibrium (LTE) so that the derived abundances become independent of the gas density.
Unfortunately, it turns out that the LTE approximation is invalid in our case. We used the ratio between the $^{13}$CO 2-1 and 1-0 lines to measure the excitation temperature unaffected by optical depth effects. It varies between 0.6 and 0.8 across our strip. The 1-0 transition is only measured at a coarser resolution than the rest of the data (see Sect.~\ref{sect_complementary}), but as all maps are relatively smooth, this still might give reasonable results. However, the ratio translates into an excitation temperature of only 5--6~K \citep{MangumShirley2015}, much lower than the typical kinetic temperature in diffuse interstellar gas or molecular clouds \citep{Tielens.book}. The value is also in contradiction to the observed peak intensity of the CO line that falls at about 8~K, corresponding to a Rayleigh-Jeans corrected excitation temperature of about 13~K in an optically thick line, thereby providing a lower limit to the actual excitation temperature. Obviously, the observed gas is not in LTE. This indicates that the gas is at densities below the critical density for the 2-1 transition of about 7000~\pccm{} \citep{Ossenkopf2000}. 

From the combination of the 1-0/2-1 ratios of the CO and $^{13}$CO transitions with the line intensities, we can instead derive all four physical parameters, the gas temperature and density and the columns of CO and $^{13}$CO, when assuming uniform beam filling of all lines.  
We use the non-LTE molecular radiative transfer code RADEX \citep{vanderTak2007} to simulate the line intensities and perform a least-squares fit to the observed mean intensities from Table~\ref{tab_integrated}. We use the velocity dispersion FWHM of 0.95~\kms{} as measured for the $^{13}$CO lines, the escape probability formalism for a spherical configuration \citep{Ossenkopf1997}, and assume that H$_2$ is the main collision partner in the gas, dominating the collisional excitation.
The fit of the intensities computes a four-dimensional $\chi^2$ distribution in the space of parameters that we define as kinetic temperature, H$_2$ density, $^{13}$CO column, and CO/$^{13}$CO abundance ratio. The minimum gives the numerically best solution. 

Figure~\ref{fig_radexfit} shows two-dimensional cuts through the four-dimensional $\chi^2$ surface measuring the match between the observed intensities and the values simulated by RADEX. All cuts are taken at the location of the $\chi^2$ minimum. The $\chi^2$ values are given on a logarithmic scale. As only values in the order of unity, this means in blue colors, indicate a valid solution, this describes a small spot in the parameter area. The parameters are well constrained, actually better than needed for a general assessment of the typical behavior within our strip. To allow for a physical understanding of the observational constraints we added four contour lines to indicate the parameter combinations where the observed values are matched. Black and white lines stand for the integrated intensities of CO 2-1 and $^{13}$CO 2-1. 
The ratios of CO 1-0/CO 2-1 and $^{13}$CO 1-0/$^{13}$CO 2-1 are shown as red and brown lines, respectively. A crossing of all lines in one point would indicate the perfect solution. The plots show that the kinetic temperature of the gas is best constrained by the CO 2-1 intensity, the gas density by the $^{13}$CO 1-0/$^{13}$CO 2-1 ratio, and the total column density and the CO/$^{13}$CO column density ratio by the CO 2-1 intensity and the CO 1-0/CO 2-1 ratio. 

The $\chi^2$-minimum is met for an H$_2$ density of 1000~\pccm{}, a kinetic gas temperature of 20.5~K, a $^{13}$CO column density of $1.6\times 10^{15}$\pscm{}, and a CO/$^{13}$CO abundance ratio of 70. At that column even $^{13}$CO turns slightly optically thick with a line-center optical depth of 0.3 and 0.7 for the 1-0 and 2-1 transition, respectively.
Using the density and temperature obtained from CO and $^{13}$CO, we can look up the mean \CI{} 1-0 intensity of 2.9~\Kkms{} in the RADEX grid and read a column of atomic carbon of $6.2\times 10^{16}$~\pscm{} and an optical depth of the 1-0 line of 0.6. We can translate the columns of CO and atomic carbon into a total gas column, using the total gas-phase carbon abundance of $X$(C)/$X$(H)$=2.34\times 10^{-4}$ \citep{SimonDiaz2011}, and obtain $N\sub{H}=4.7\times 10^{20}$~\pscm{} for the gas traced by CO and $N\sub{H}=2.6\times 10^{20}$~\pscm{} for the gas traced by \CI{}. This is only about 20\,\% of the total column as measured through the dust emission. About the same amount of gas can be ``hidden'' in ionized carbon within the upper limit of the \CII{} line intensity of 0.2~\Kkms{} when assuming the same excitation conditions, but that still leaves about half of the total gas column unexplained. As the ratios between the lines and the dust-based column density vary only weakly along the strip, this ratio is independent of the exact location in the strip.

\begin{figure*}
   \centering
   \includegraphics[angle=0,height=4.1cm]{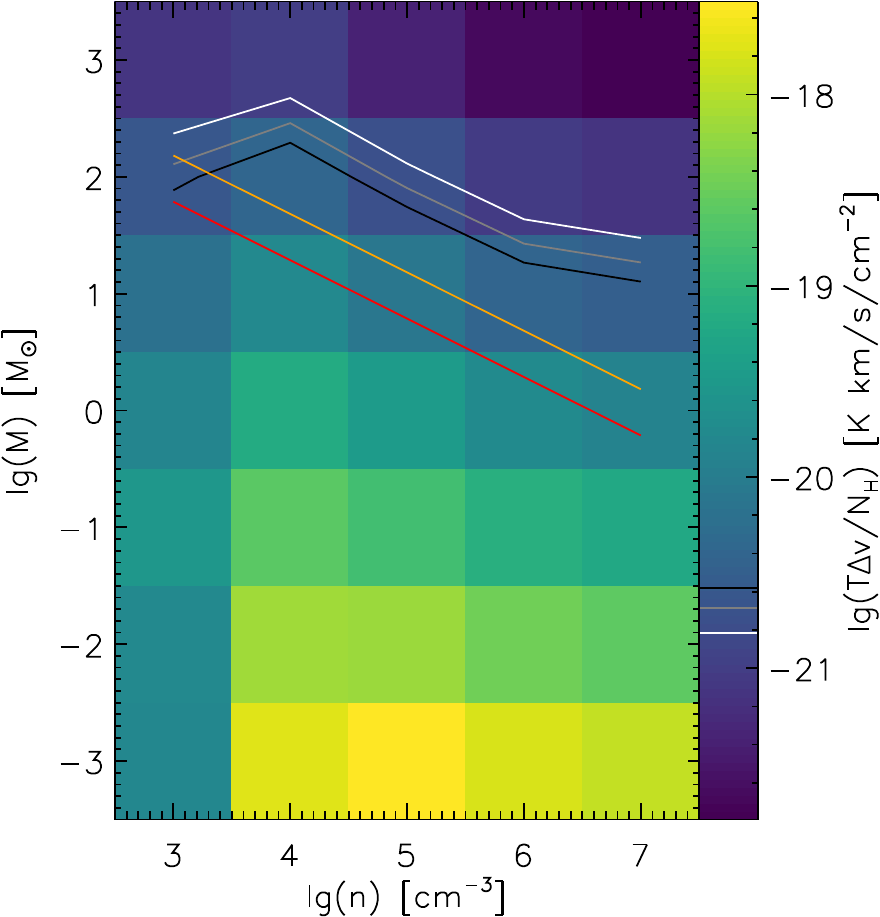}
   \includegraphics[angle=0,height=4.1cm]{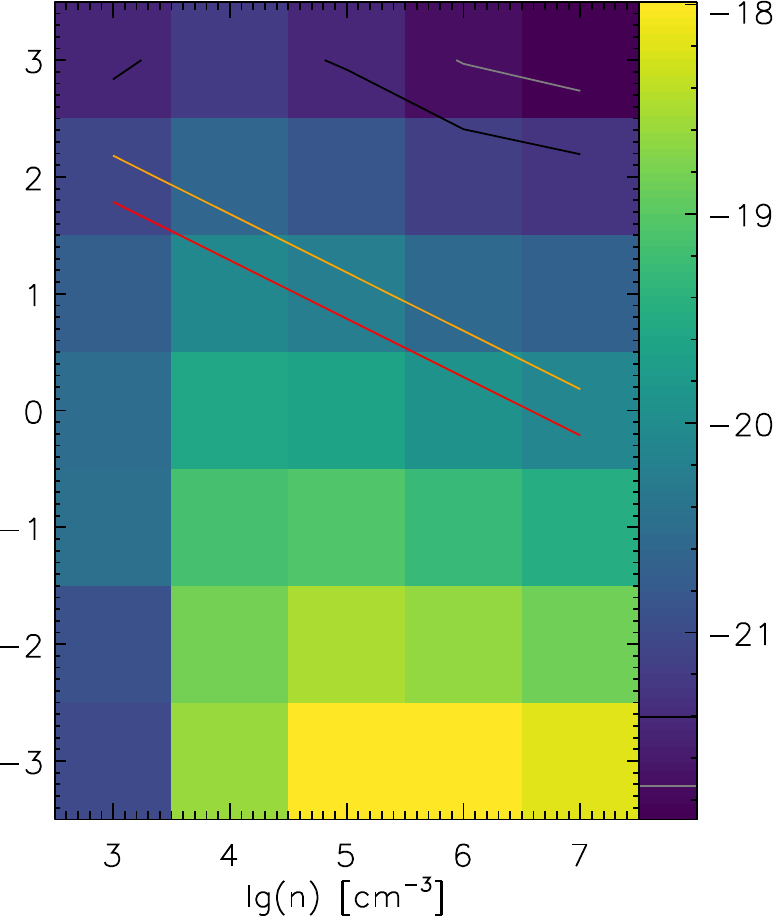}
   \includegraphics[angle=0,height=4.1cm]{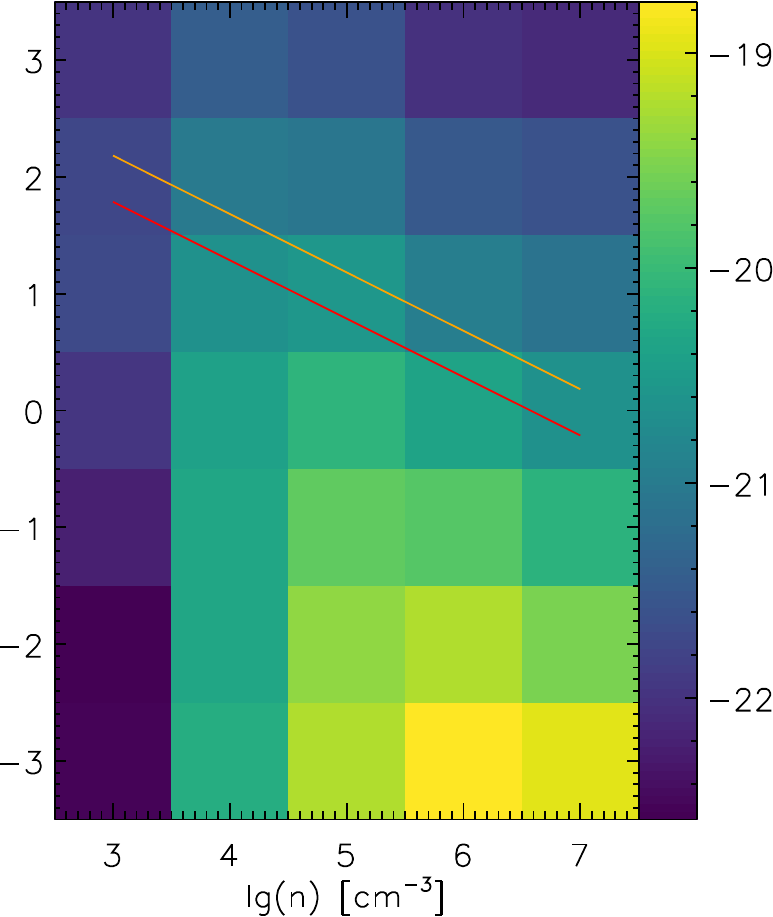}
   \includegraphics[angle=0,height=4.1cm]{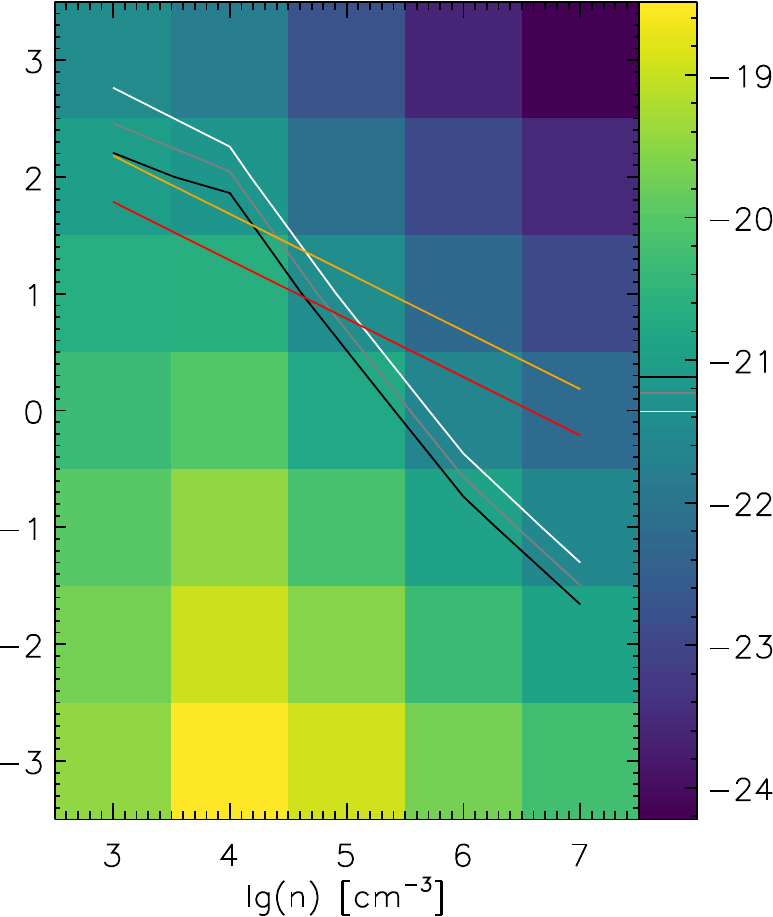}
   \includegraphics[angle=0,height=4.1cm]{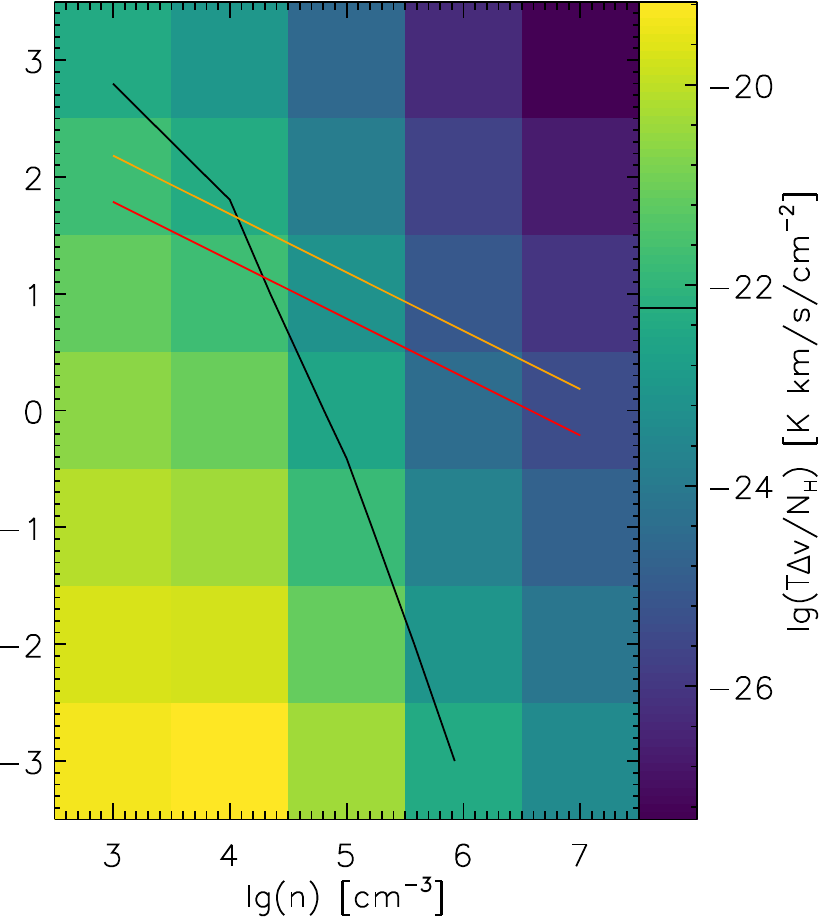}
   \caption{Integrated line intensities predicted by the KOSMA-$\tau$ PDR model relative to the column density of the model clumps. The panels from left to right show CO 2-1, $^{13}$CO 2-1, C$^{18}$O 2-1, \CI{} 1-0 and \CII{}. The white, grey, and black lines indicate the minimum, mean and maximum of the observed ratio. For \CII{} we only have an upper limit, for C$^{18}$O 2-1 even the maximum falls outside of the covered parameter range, being lower than the displayed values. The red and orange contours indicate the lowest and highest column density in the region obtained from the dust observations.}
    \label{fig_model_relative_to_nh}%
\end{figure*}

\begin{figure*}
   \centering
   \includegraphics[angle=0,height=4.1cm]{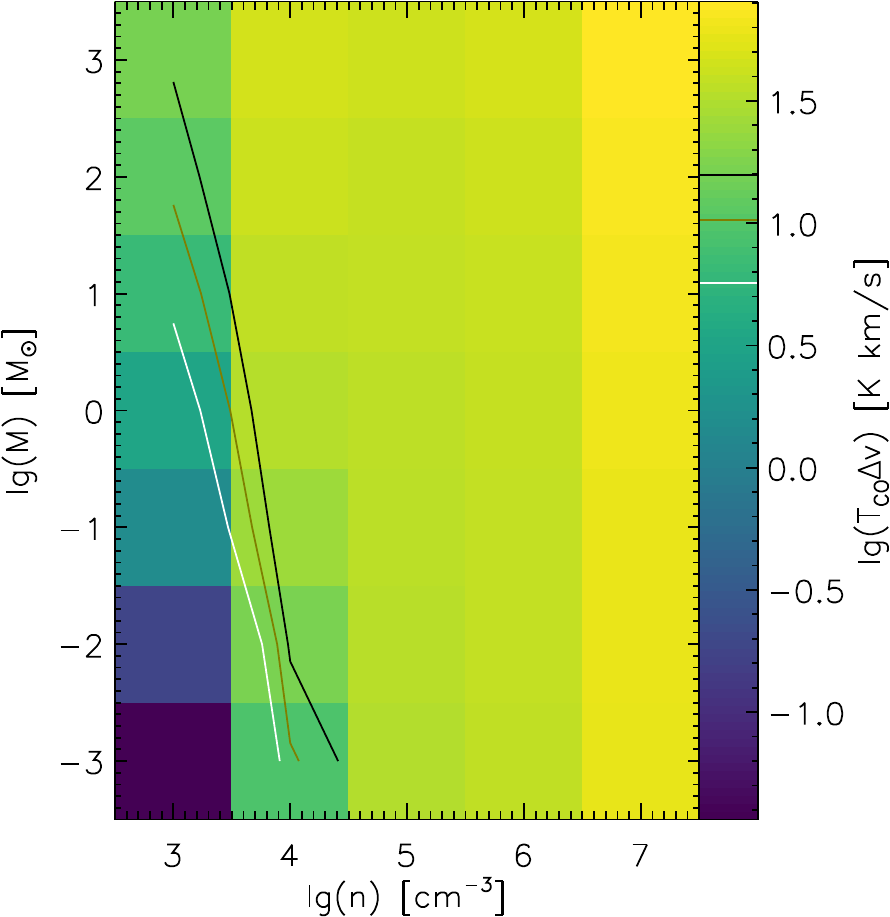}
   \includegraphics[angle=0,height=4.1cm]{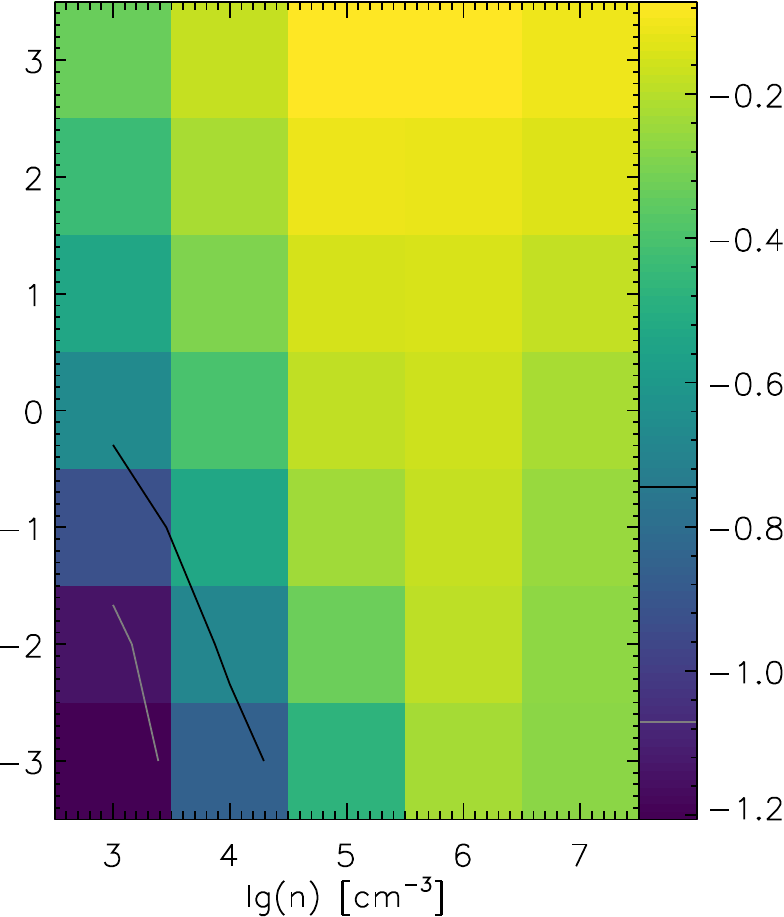}
   \includegraphics[angle=0,height=4.1cm]{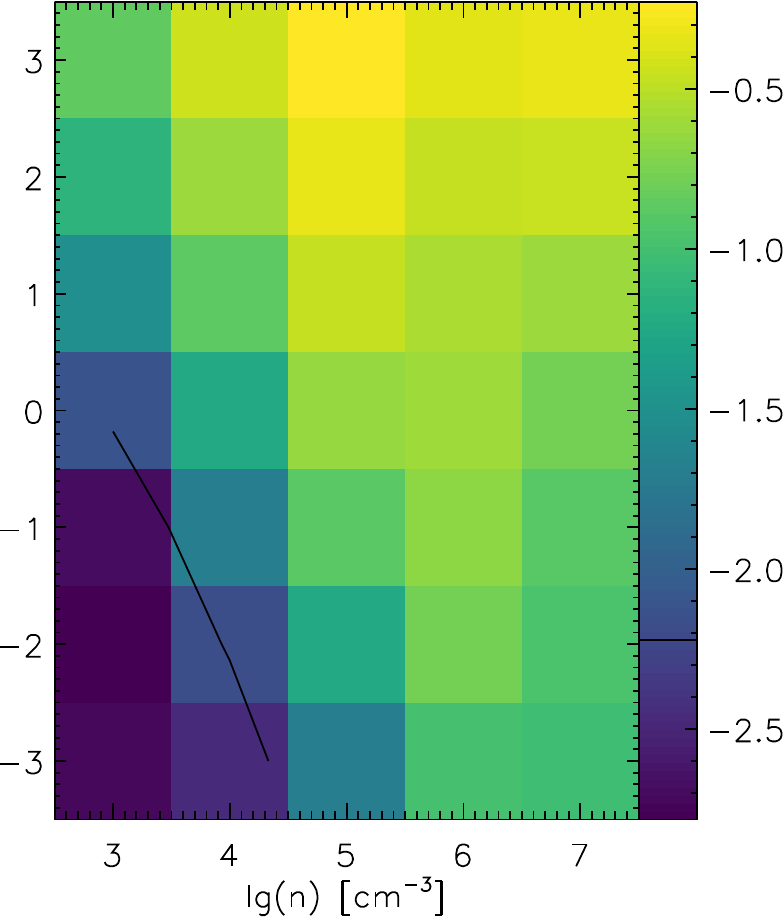}
   \includegraphics[angle=0,height=4.1cm]{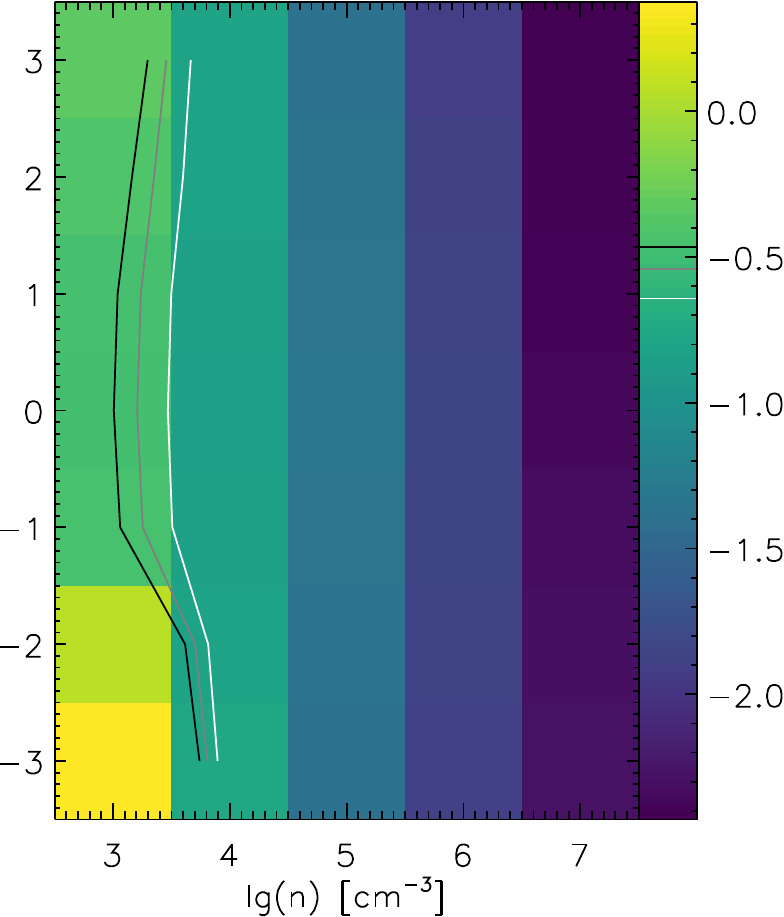}
   \includegraphics[angle=0,height=4.1cm]{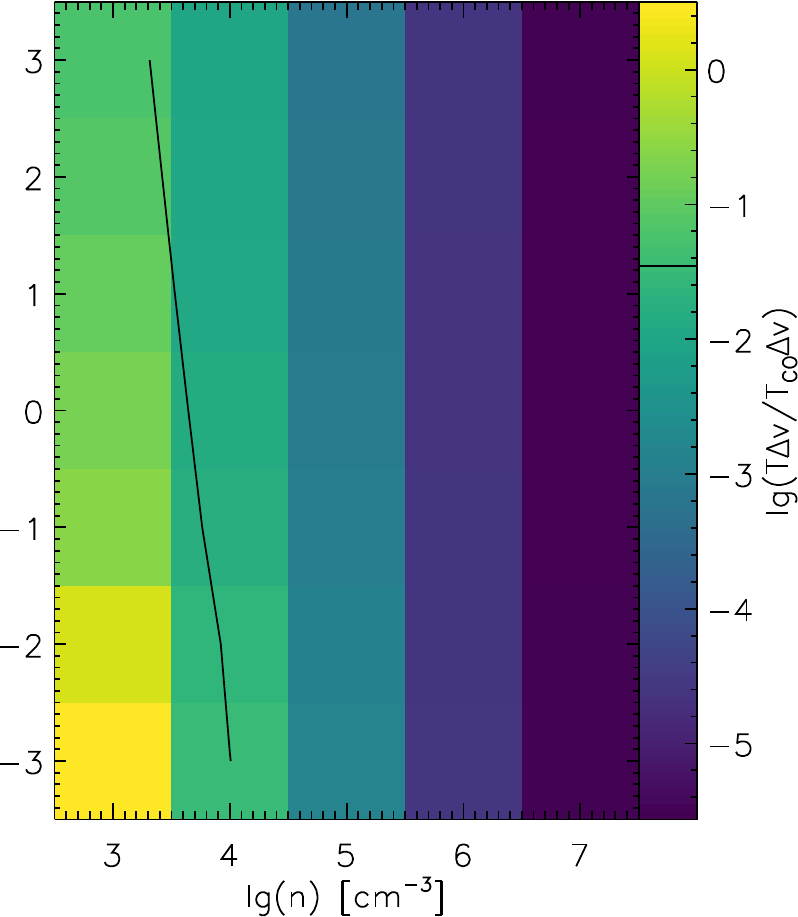}
   \caption{Integrated line intensities predicted by the KOSMA-$\tau$ PDR model relative to the CO 2-1 intensity. The left panel shows this CO 2-1 intensity, the subsequent panels give the ratios for $^{13}$CO 2-1, C$^{18}$O 2-1, \CI{} 1-0 and \CII{}. The white, grey, and black lines indicate the minimum, mean and maximum of the observed ratios.}
    \label{fig_model_relative_to_co}%
\end{figure*}

\subsection{PDR modelling}
\label{sect_pdrmodelling}

Instead of deriving individual abundances of the different species, we can compare the observed intensities with a chemical PDR model that self-consistently predicts all abundances and excitation temperatures as a function of fewer parameters. A number of PDR models is available for this type of astrochemical modelling of the region \citep[see e.g.][]{Roellig2007, Wolfire2022}. Most of them assume a one-dimensional infinite slab in a plane-parallel geometry. Instead, we used the KOSMA-$\tau$ PDR model \citep{Roellig2013, Roellig2022} that tries to represent the large importance of surfaces by switching to a spherical configuration or a superposition of many spherical clumps, thus exhibiting a larger surface to volume ratio than the plane-parallel models. In this way we should better match the observed large \CI{}/CO intensity ratio in our region that indicates a large fraction of dissociated gas in UV illuminated surface layers. Moreover, KOSMA-$\tau$ includes the full isotopic chemistry allowing us to also compare the abundances of $^{13}$CO and C$^{18}$O. The model assumes a radial density gradient with a central density that is about ten times larger than the surface density that we give as parameter in the figures here.

We do not perform a combined $\chi^2$ fit of all observed ratios but rather inspect the behavior of the different ratios in a model grid spanning a range of surface densities and clump masses to constrain under which conditions the observed ratios are reproduced. As there are no bright UV sources in the vicinity of our region, we stick to the normal impinging UV flux measured in the solar neighborhood of one Draine field, \citep[$\chi\sub{D}=8.96\times 10^{-14}$~erg~cm$^{-3}$, ][]{DraineBertoldi1996}. In Appx.~\ref{appx_pdr} we show the corresponding results for a ten times higher UV flux,  proving that none of our conclusions would change if that assumption is violated. The model grid is based on the UdfA 2012 chemical network \citep{McElroy2013} without ice formation. All parameters are listed in \citet{AndreeLabsch2017}.

Figure~\ref{fig_model_relative_to_nh} shows the model predictions for the ratio of the five integrated line intensities relative to the column density when varying the clump density and mass. By considering these ratios the results are independent of variations of the total column density in the beam or a non-uniform beam-filling, but focus on the line emissivity of the observed material. Yellow patches correspond to higher line intensities from the same column of material, dark-blue patches to low emissivities. We see for all of our lines that very dense and massive clumps produce only very low line emissivities, because those clumps would be cold and optically thick. The strongest line emissivity is reached for the smallest clumps of only $10^{-3}~M_\sun$, but the density of the peak emissivity varies strongly between the five lines. For C$^{18}$O 2-1 we obtain the brightest line per column for a clump surface density of $10^6$~\pccm{}, for \CII{} at $10^3$~\pccm{}.

The plots contain two types of contours. The white, grey and black contours represent minimum, mean and maximum of the observed ratios from Table~\ref{tab_integrated}. Not all subplots show all three contours. For $^{13}$CO 2-1 the observed minimum is not reached within the scanned parameter range and even the mean and maximum is only predicted for the densest and most massive clumps. For C$^{18}$O 2-1, even the maximum observed value falls below all model predictions so that no contours are shown. The lowest model predictions, thereby being closest to the observed range, occur for the small and thin clumps in the lower left corner of the parameter range where almost all C$^{18}$O is photo-dissociated but some CO and $^{13}$CO exist due to their higher self-shielding. For \CII{} that is the parameter range with the highest predicted intensity. However, the \CII{} upper limit from the possible tentative detection, indicated by the black line, excludes all the high emissivities predicted by the model in the left lower part of the parameter range. 

The second type of contours, the red and orange lines, indicate model parameters where the column densities observed in the strip match the column densities of the individual model clumps. This is not a strict constraint on the models. In parameter regions where the model column density falls below the observed column density (lower left corner), the observed column might be the sum of multiple clumps along the line of sight at the same velocity. If the model clump column density exceeds the observed column (upper right corner) only part of the clumps might fall into the beam so that the measured column density within the telescope beam falls below the clump column. This makes sense for the massive low-density clumps in the upper left corner of the parameter range where the physical size of the model clumps exceeds the beam width at 1.0~kpc distance. However, the narrow line width and the smooth distribution of the emission across the map, even at the best resolution, indicate that both effects probably play only a minor role so that the region selected by the two lines is at least an approximate guideline.

Figure~\ref{fig_model_relative_to_co} shows again model ratios of the different integrated line intensities, but this time relative to the CO 2-1 intensity. In this way we characterize the CO-bright molecular gas only. The CO 2-1 intensity itself is shown in the left panel. The relatively low absolute intensity of CO 2-1 indicates gas at moderate densities of 10$^3$-10$^4$~\pccm{}, but we see only a very weak sensitivity to the clump mass. The \CI{} 1-0/CO 2-1 ratio suggests the same low densities independent of the clump mass. The measured low ratios of $^{13}$CO 2-1/CO 2-1 and C$^{18}$O 2-1/CO 2-1 further constrain the parameter range by also asking for low densities, but additionally excluding all massive clumps. This is in contradiction to the small measured \CII{}/CO 2-1 ratio that excludes all low density models, requiring densities of at least 10$^4$~\pccm{}.  

Combining the information we have to conclude that all models are invalid for our region. 
When considering only CO, \CI{}, and the total gas column, models with low density and masses around $100\,M_\sun$ would fit the data. In contrast both $^{13}$CO 2-1 and C$^{18}$O 2-1 require models where the molecules are dissociated, consisting mainly of atomic gas, They are simulated by models with low mass and density. In contrast, the low \CII{} intensity observed in the field is only consistent with mainly molecular gas, requiring high densities. Models for a higher UV field (Appendix~\ref{appx_pdr}) show that this incompatibility is independent of the UV field; a higher UV field only shifts all lines for $^{13}$CO 2-1, C$^{18}$O 2-1, and \CII{} towards larger clump masses and densities but creates no overlap. 

These tests stress that a blind $\chi^2$-fit of the observational data by the models would be rather misleading. It would hide the fundamental mismatch. It is also easy to see why switching to a plane-parallel model does not solve the problem: lowering the surface relative to the volume decreases photo-dissociation of $^{13}$CO 2-1 and C$^{18}$O 2-1 so that it would drive their line ratio relative to CO 2-1 further up, creating an even larger mismatch to the observations. 

In contrast, it might be possible to construct a PDR model that simultaneously fits the $^{13}$CO 2-1, C$^{18}$O 2-1, and \CII{} observations by using a much lower UV field, resulting in very cold clouds. A very low UV field due to some pre-shielding gas was already proposed as a solution to PDR model mismatches by \citet{Bensch2003}, but as discussed in \citet{Schneider2024}, it is very unlikely to find a UV field much below the Draine field in the Galactic plane. Hence, we have not gone into the effort of creating models for such conditions. This can be a possible approach for follow-up investigations.

\section{Discussion}
\label{sect_discussion}

Qualitatively, the intensity profiles of the CO isotopologues follow the expected sequence, where isotope-selective photodissociation due to different columns of self-shielding molecules  lead to the sequence of CO formation at relatively low column densities of attenuating dust, $^{13}$CO formation at larger columns, followed by C$^{18}$O formation at even larger columns. The correlation analysis in Sect.~\ref{sect_ratios} indicates a critical visual extinction of $A\sub{V}$=$1^m.2$ for the formation of CO, $A\sub{V}$=$1^m.9$ for the formation of $^{13}$CO, and $A\sub{V}$=$2^m.3$ for C$^{18}$O\footnote{We use the standard conversion between visual extinction and hydrogen column of $N\sub{H}/A\sub{V} = 1.87 \times 10^{21}$ cm$^{-2}$ mag$^{-1}$ \citep{Bohlin1978}.}. Figure~\ref{fig_integrated} shows that the onsets of emission from $^{13}$CO and C$^{18}$O fall within the observed strip. However, when comparing those numbers with the predictions from the model grids introduced in Sect.~\ref{sect_pdrmodelling} there are no good matches. Depending on the model parameters, the critical columns for the formation of the CO isotopologues vary slightly but they all fall in the range around $A\sub{V}$=$1^m.0$, and the spread between the three isotopologues covers less than $A\sub{V}$=$0^m.3$. Hence, the effective column shielding the gas against the UV radiation must be significantly lower than the column measured along the line of sight. This can be explained either by a significant contribution of unrelated diffuse interstellar gas to the measured dust column (see below) or by the typical geometric difference between the effective UV shielding column, $A\sub{V,eff}$ in a turbulent three-dimensional configuration and the line-of-sight column, $A\sub{V}$. \citet{Seifried2020} found $A\sub{V,eff} \approx 1/3 A\sub{V}$ for a set of hydrodynamic and magneto-hydrodynamic simulations of the turbulent ISM within the SILCC-Zoom project \citep{Walch2015,Seifried2017}. However, in both cases we still cannot explain why the spread between the isotopologues in our observations is much larger than in the models.

Our radiative transfer analysis showed a discrepancy between the gas column obtained from the different carbon-bearing species and the column derived from the dust emission when assuming uniform excitation conditions in the gas. Counting the carbon in the gas phase provided less than half of the column measured through the dust emission. Consequently, our methodical assumptions need to be questioned. As discussed above there is, of course, no chemically homogeneous configuration. Instead we should see a layered structure with ionized carbon at the cloud surface only weakly shielded from the interstellar radiation field, then atomic carbon and CO at intermediate depths, then $^{13}$CO, and finally C$^{18}$O deepest within the clouds, typically also associated with largest densities.
This is consistent with the fact that the measured CO/$^{13}$CO abundance ratio of 70 is slightly higher than the value obtained from \citet{LangerPenzias1993} and \citet{Wilson1999} for a Galactocentric radius of 9~kpc for the 14~\kms{}-component (see Sect.~\ref{sect_results}). However, when inspecting the solution in Fig.~\ref{fig_radexfit} for the impact of a density variation between $^{13}$CO and CO, the dependence is very weak. The conclusion is even stronger for the \CI{} 1-0 line. Its emissivity per carbon column is quite constant over a large range of temperatures and densities. 
\citet{Papadopoulos2022} have shown that for a wide range of  ISM conditions ($n_{\rm  H_2} = 100-10^{4}$~\pccm{} and $T_{\rm  kin} = 20-80\,{\rm
K}$) the \CI{} 1-0 emissivity changes only by 20\,\%. 
The only explanation for the discrepancy in the frame of chemical differentiation of the region is the existence of some diffuse envelope or line-of-sight contamination with densities as low as 10~\pccm{}. In this gas all carbon would be in the form of C$^{+}$ but its collisional excitation is too low to make it visible in the \CII{} line \citep{Goldsmith2012}. In this scenario, only half of the column density that we see in the dust emission would be associated to the gas that we observed, while the other half stems from the diffuse medium. Unfortunately, this explanation is in contradiction to the low dust temperature determined in the region. The PPMAP analysis \citep{Marsh2017} shows most of the dust at temperatures around 15~K, much colder than dust in diffuse clouds.

This leads, however, to an alternative explanation: gas freeze-out. The gas phase abundances used here were calibrated for the Orion Molecular Cloud \citep[OMC1,][]{SimonDiaz2011}. Dust temperatures of 15~K are significantly colder than in active star forming regions like OMC1 so that we might assume that half of the carbon from the gas phase is frozen to the grains. This is consistent with the temperature dependence of the C$^{18}$O depletion factor measured by \citet{Lewis2021} in the California Cloud. A resulting gas phase abundance as low as $X(\mathrm{C})/X(\mathrm{H})=1.2\times 10^{-4}$ would provide an alternative explanation for the discrepancy. However, \citet{Lewis2021} found a stronger depletion for $^{13}$CO while we see a significant under-abundance of  C$^{18}$O. Moreover, PDR models predict freeze-out only deep in the clouds, not close to the surface where C$^{18}$O just starts to form. We have to conclude that both explanations are so far questionable.

\citet{Benedettini2020} compared the CO 1-0 and $^{13}$CO 1-0 intensities with the dust based column densities and obtained conversion factors between CO intensity and H$_2$ column corresponding to CO 1-0$/N\sub{H}=1.5 \times 10^{-21}~\Kkms{}\mathrm{ cm^{2}}$ and $^{13}$CO 1–0$/N\sub{H}=4.2 \times 10^{-22}~\Kkms{}\mathrm{ cm^{2}}$ for the FQS. Our values of $2.6 \times 10^{-21}~\Kkms{}\mathrm{ cm^{2}}$ and $3.2 \times 10^{-22}~\Kkms{}\mathrm{ cm^{2}}$ for the two ratios (Table~\ref{tab_integrated}) deviate in opposite directions, indicating that our field is not representative for the whole region. The strip is relatively CO-bright, compared to the whole FQS area, while it is still only weakly shielded so that $^{13}$CO is less abundant. 
If the good correlation between \CI{} 1-0 and the CO transitions holds for the full area of the FQS this would suggest a significant and extended \CI{} brightness from that part of the Galaxy.

In contrast to many previous observations, we observed a relatively high intensity of the \CI{} line when compared to CO. Our fits of the line ratios provided a \CI{} 1-0/CO 2-1 ratio (in K~\kms{}) of 0.27 and a \CI{} 1-0/$^{13}$CO 2-1 ratio of 1.6 for the 14~\kms{} component. For the 20~\kms{} component, the \CI{} 1-0/CO 2-1 ratio even went up to 2.6. 
For the Orion A cloud, \citet{Arunachalam2023} and \citet{Labkhandifar2023} found typical ratios for \CI{} 1-0/$^{13}$CO 2-1 of about 0.3 and for \CI{} 1-0/CO 2-1 of about 0.05. \citet{Lee2022} measured for the ATLASGAL clumps a mean ratio \CI{} 1-0/$^{13}$CO 2-1 of 0.56 and a tendency towards higher ratios for less evolved clumps. Their sample of 35 diffuse clumps shows a mean ratio of 1.0. Aligning our results with those data we see a clear trend from low ratios between \CI{} 1-0 and the molecular lines for dense evolved regions, harboring active star formation, to high ratios for diffuse clouds at low radiation fields. In this sequence, our observations currently provide the low density end with the highest ratios.

This is in line with unresolved observations of whole galaxies. \citet{Topkaras2024} determined for a large number of extragalactic observations, integrating over whole galaxies, a typical ratio between the integrated intensities \CI{} 1-0/CO 1-0 $\approx$ 0.22 or \CI{} 1-0/CO 2-1 $\approx$ 0.26, respectively. This is five times higher than the Galactic value for the bright regions but in line with our number for the 14~\kms{} component and below the ratio measured in the 20~\kms{} component. When integrating over a whole galaxy that always contains a mixture of dense and bright regions with diffuse gas, the atomic carbon is predominantly seen from the diffuse regions.
The \CI{} emission from galaxies is dominated by material at moderate densities around $10^3$~cm$^{-3}$, not by the small bright cores. In that sense, it seems worth looking back at low-resolution single dish mapping observations of the Milky Way and nearby galaxies, not focusing on the new capabilities of modern interferometers only. 

Future observations should examine the 20~\kms{} component where the \CI{} emission exceeds that of CO. One can speculate that this indicates a large fraction of CO-dark molecular gas in that component. Unfortunately, the GOTC+ observations did not cover the component and high-resolution \HI{} absorption are neither available yet.
A large fraction of the CO-dark molecular gas may be traced rather by \CI{} emission than by \CII{} \citep[see][]{Pineda2013}. As \CII{} based investigations already showed fractions of CO-dark molecular gas between 20\,\% and 80\,\% \citep[e.g.][]{Langer2014,Xu2016,Chen2015}, the actual amount of molecular material that is not traced by the standard CO observations may be significantly higher. Combining the existing knowledge on an increasing fraction of CO-dark molecular gas with increasing Galactic latitude \citep{Luo2024,Xu2016} with our indications of CO-dark molecular gas even in the Galactic midplane traced through \CI{} emission suggests that we may still miss a significant fraction of the Milky Way's ISM in all observations done until today.
This is a strong justification for the science case of the GEco survey mapping large areas in the Galaxy.


\section{Conclusions}
\label{sect_conclusions}

When mapping regions of the the Milky Way that are not prominent star-formation sites, several of the relations calibrated for the massive star-forming regions break down. By selecting such an inconspicuous region for a detailed mapping in \CI{} 1-0 and the different CO isotopologues we found that the gas fraction detected in CO is rather low. A significant fraction of the gas is traced through \CI{}, leading to a \CI{} 1-0/CO 2-1 ratio that can be as high as 2.6 for our 20~\kms{} component. This could be indicative of CO-dark molecular gas, but a clear assignment requires better \HI{} data. Unfortunately, the sum of the carbon-bearing gas seen in the CO isotopologues, \CI{}, and ionized carbon, visible through \CII{} emission, falls short of the column measured by dust emission by about a factor of two, if we assume a gas density around $10^3$~\pccm{} that is consistent with the observations of \CI{}, CO and $^{13}$CO. 

The optical depth of the \CI{} 1-0 line is similar to that of the $^{13}$CO lines, but the spatial distribution is very different due to the layering structure of PDRs. In contrast, we found a very close correlation to the intensity distribution of the CO lines and a match to their line width, in spite of a much larger optical depth of the CO lines. This indicates that neither CO nor $^{13}$CO follow the abundance of atomic carbon in translucent clouds. Atomic carbon must be more extended both spatially and in velocity space. Large-scale mapping of \CI{} in the Milky Way, including the non-prominent, but statistically important diffuse regions, is therefore essential for deriving the correct distribution of atomic carbon in our Galaxy. This is the goal of the planned GEco project at CCAT/FYST, that will map the Milky Way in \CI{}. 

Qualitatively, our observations seem to confirm the astrochemical models predicting a layering of species in PDRs. They show the expected sequence
with extended \CI{} and CO, $^{13}$CO and C$^{18}$O above different critical column densities, governing the self-shielding of the molecules. Quantitatively, however, we note significant mismatches. The models predict a smaller separation between the onsets of CO, $^{13}$CO and C$^{18}$O emission and they predict either more \CII{} emission than observed -- in case of diffuse gas -- or more C$^{18}$O emission than observed -- in case of dense gas. This hints towards some fundamental deficiency in the current PDR modelling or some, so-far overlooked, mechanism of additional UV attenuation.

It is, of course, impossible to draw any statistical conclusion from the tiny strip observed here. It fits, however, into a consistent picture for the \CI{} emission from galaxies where the emission integrated over a whole galaxy represents a mixture of bright and diffuse regions. Bright regions, that were the focus of most previous studies, are only responsible for a small contribution to the total \CI{} emission while more diffuse gas, as observed here, actually dominates the sum.

\begin{acknowledgements}
We thank Arnaud Belloche for help with the preparation of the observations and their execution. We thank Markus R\"ollig for many useful discussions.
This work is supported by the Collaborative Research Center 1601 (SFB 1601 sub-projects A1, A6, and B2) funded by the Deutsche Forschungsgemeinschaft (DFG, German Research Foundation) – 500700252.  
\end{acknowledgements}

\bibliographystyle{aa}
\bibliography{refs} 

\begin{appendix}

\section{\CI{} at the full spatial resolution}
\label{appx_highres}

\begin{figure}
   \centering
\includegraphics[angle=0,height=8.1cm]{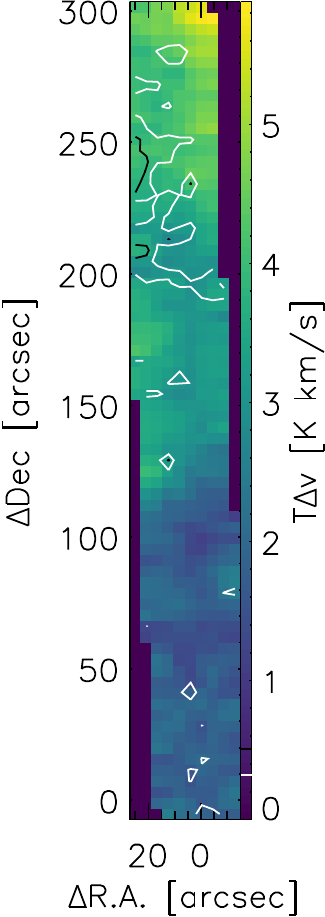}
\hspace*{1cm}   \includegraphics[angle=0,height=8.1cm]{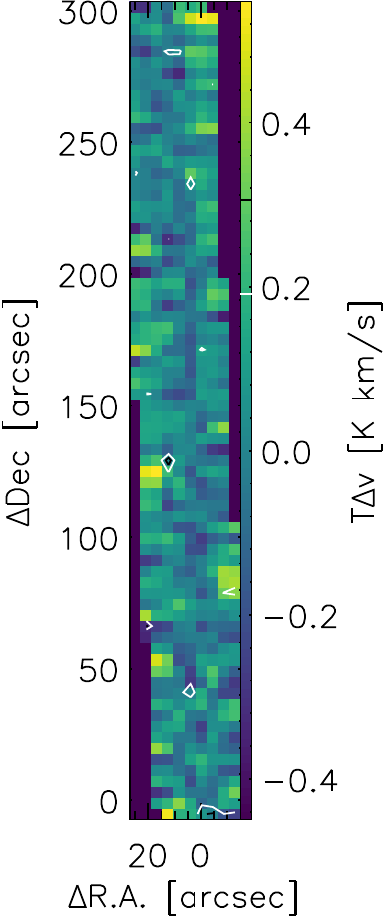}
   \caption{Integrated \CI{} map at a resolution of 14.3~arcsec (left) and difference to the low-resolution map from Fig.~\ref{fig_maps} (right). Colors give the integrated intensity for the 14~\kms{} component, contours show the 20~\kms{}-component.}
    \label{fig_highres}
\end{figure}

The APEX resolution of 13~arcsec at 492~GHz allows us to study spatial variations at smaller scales than for the other tracers. Figure~\ref{fig_highres} (left) shows the integrated line map, equivalent to Fig.~\ref{fig_maps}, at a spatial resolution of 14.3~arcsec. Some small scale variations are visible, but they can be partly due to the noise in the data, not tracing actual variations. Therefore the right hand side shows a plot that gives the difference map between this 14.4~arcsec map and the map smoothed to the 29.7~arcsec of Fig.~\ref{fig_maps}. As the noise level of the integrated lines is 0.2~\Kkms{} we can conclude that most of the spatial variations can just be due to noise. Overall the high resolution data are consistent with the smooth \CI{} distribution discussed in Sect.~\ref{sect_results}.
\vfill\break

\onecolumn

\begin{figure*}[ht]
   \centering
   \includegraphics[angle=0,height=4.1cm]{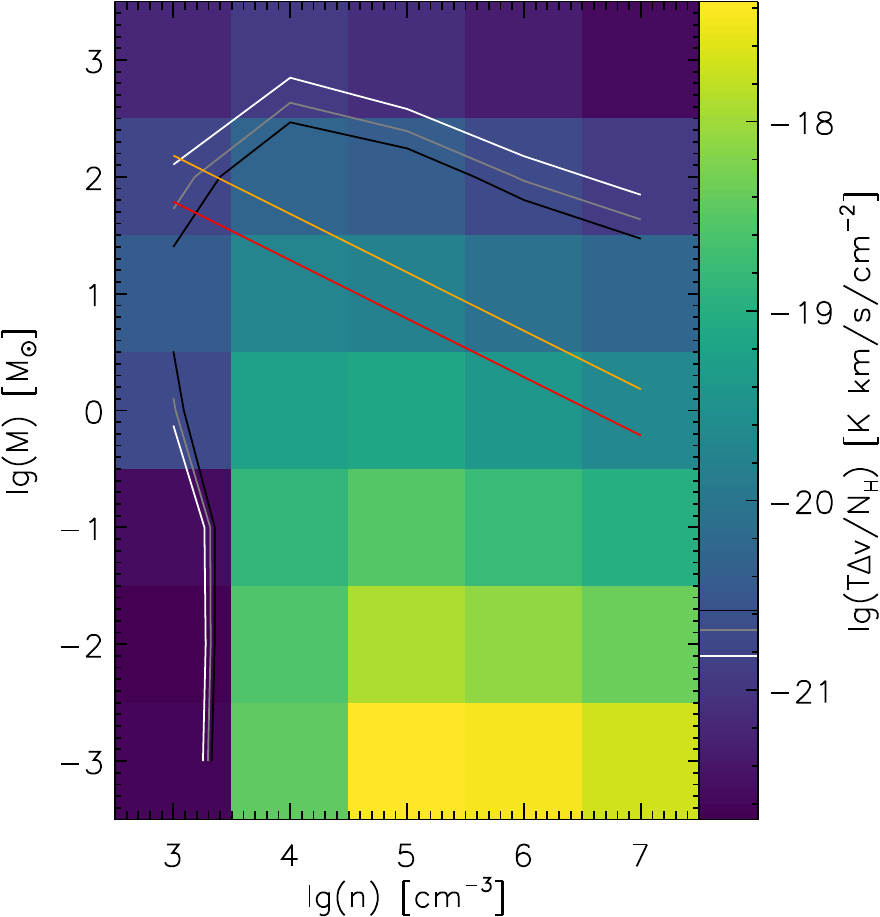}
   \includegraphics[angle=0,height=4.1cm]{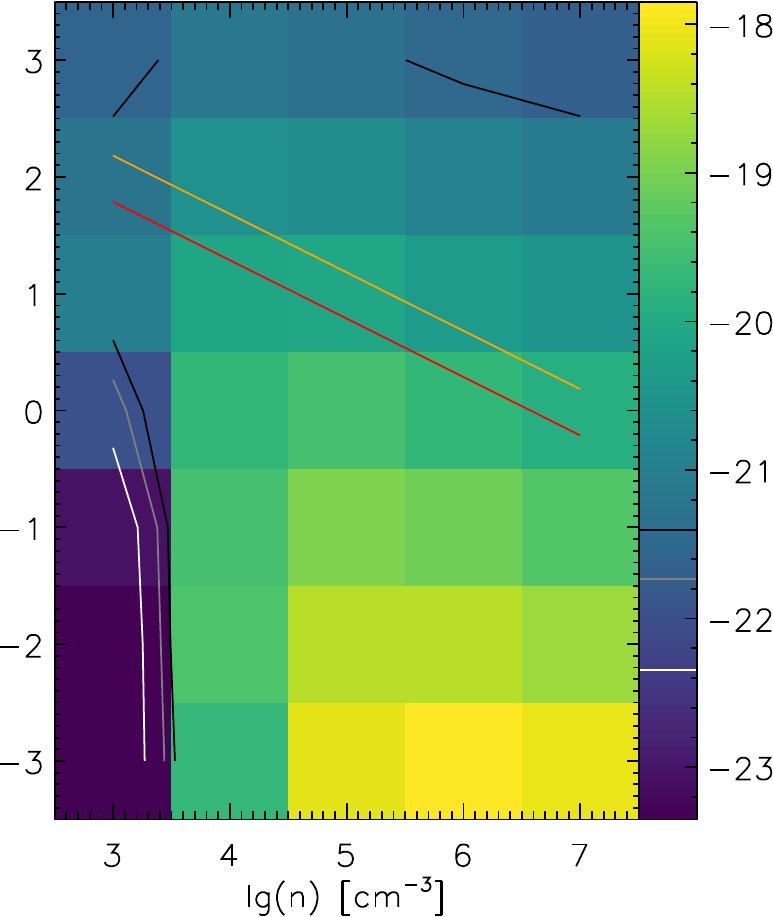}
   \includegraphics[angle=0,height=4.1cm]{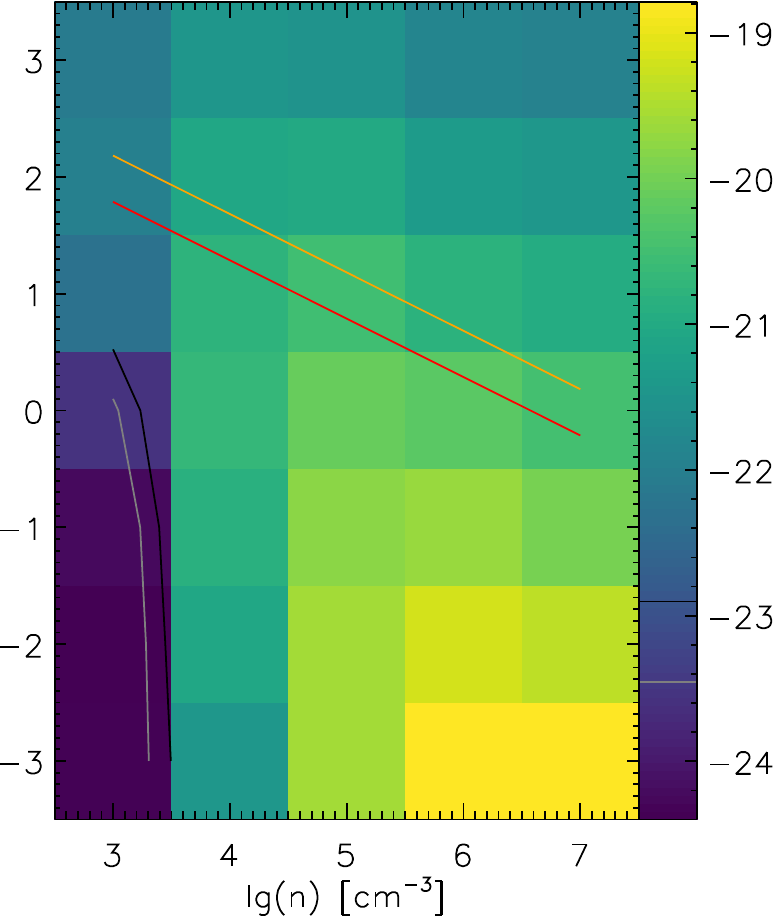}
   \includegraphics[angle=0,height=4.1cm]{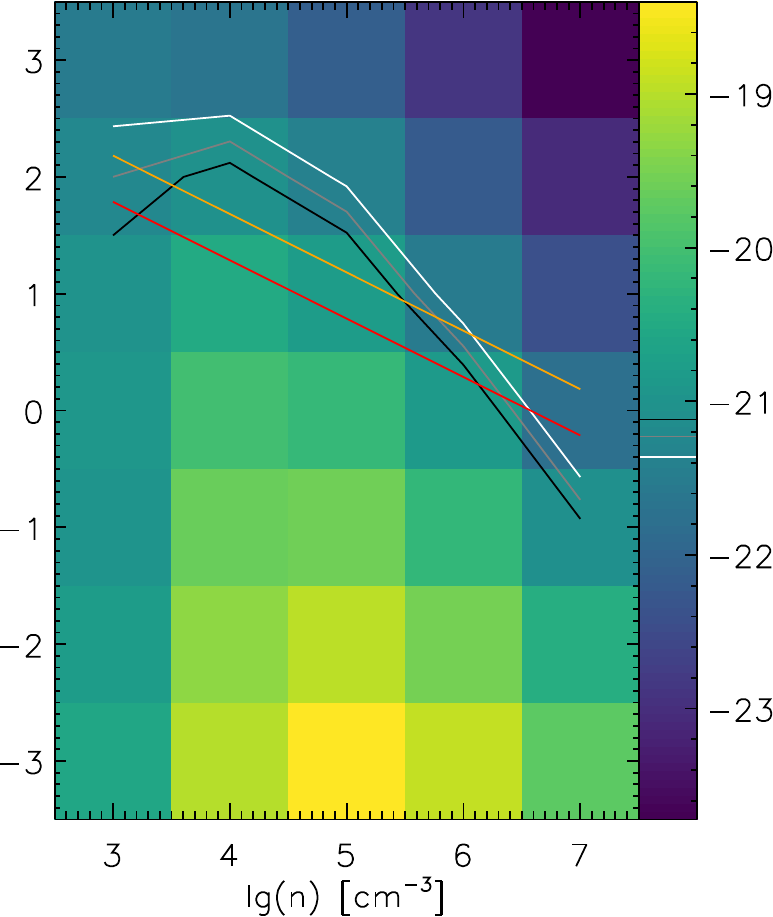}
   \includegraphics[angle=0,height=4.1cm]{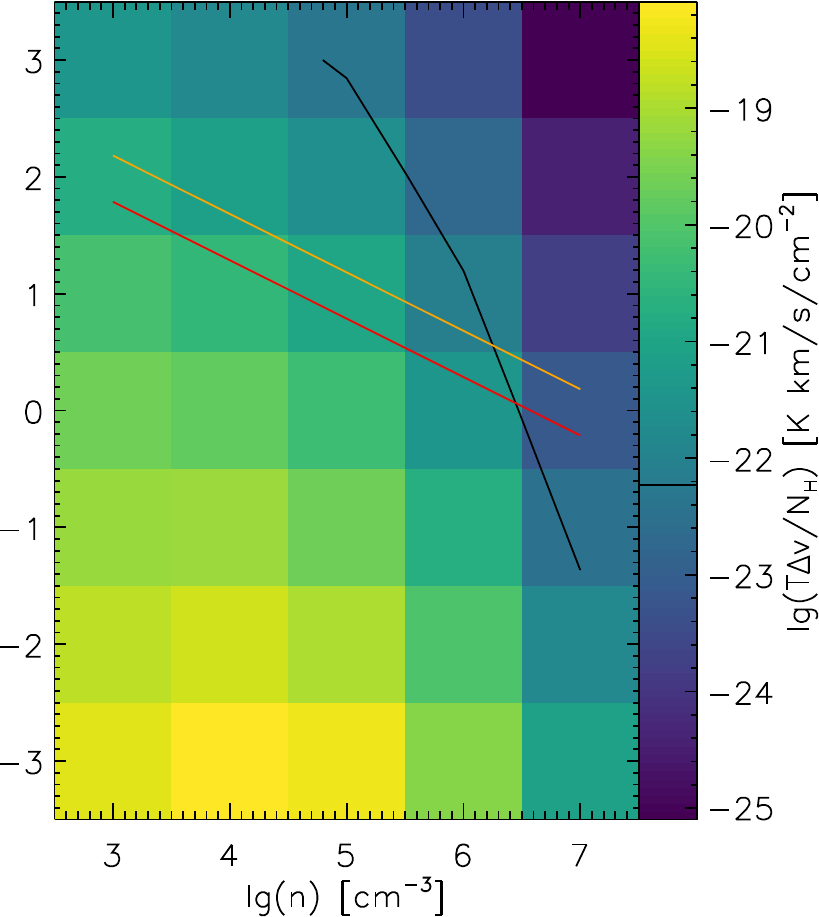}
   \caption{Same as Fig.~\ref{fig_model_relative_to_nh} but for a UV radiation field of 10$\chi_0$.}
    \label{fig_model_relative_to_nh_chi10}%
\end{figure*}

\begin{figure*}[ht]
   \centering
   \includegraphics[angle=0,height=4.1cm]{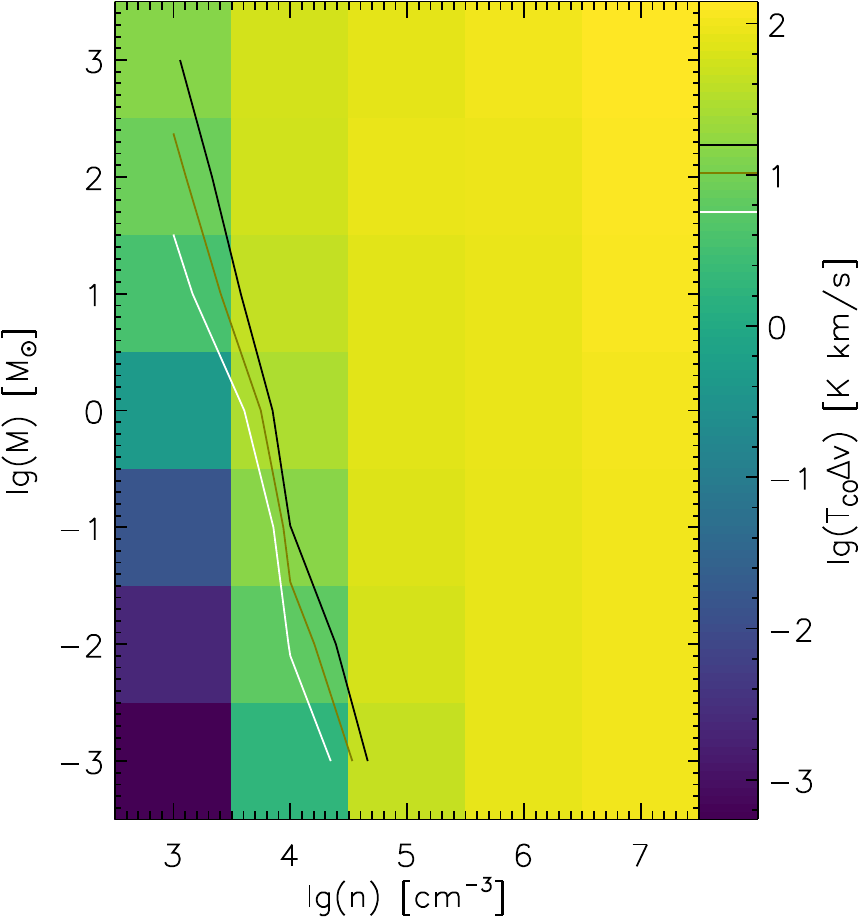}
   \includegraphics[angle=0,height=4.1cm]{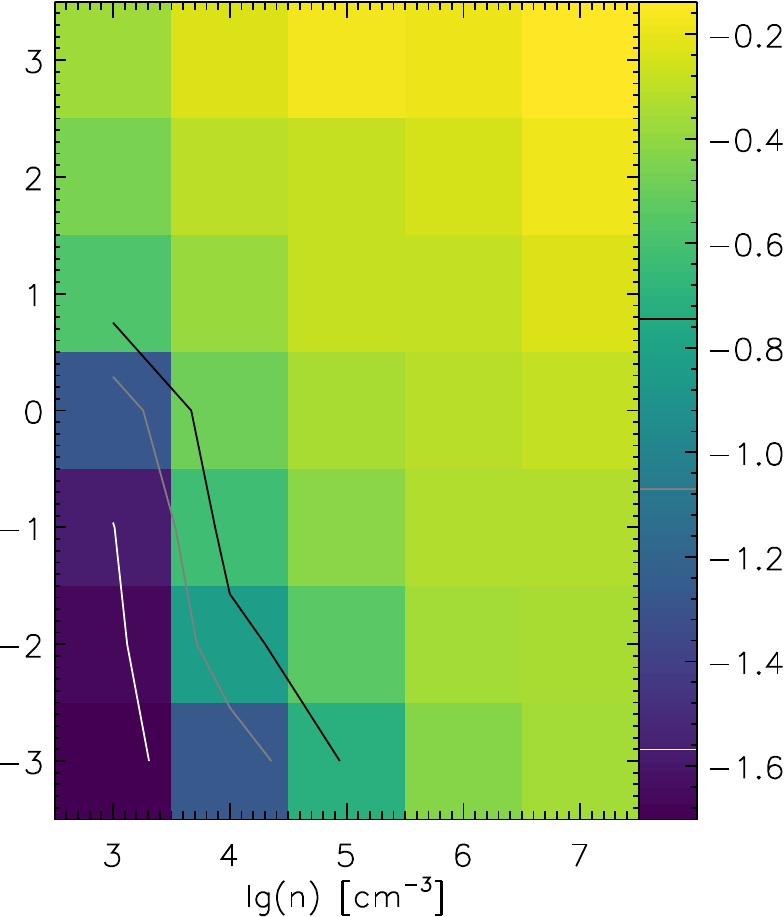}
   \includegraphics[angle=0,height=4.1cm]{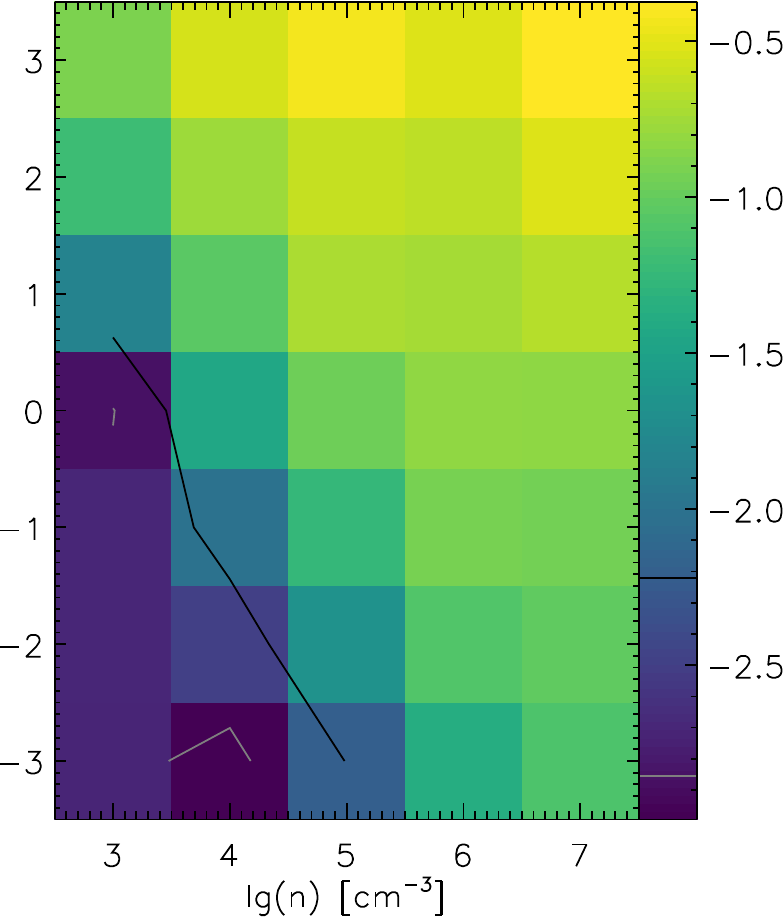}
   \includegraphics[angle=0,height=4.1cm]{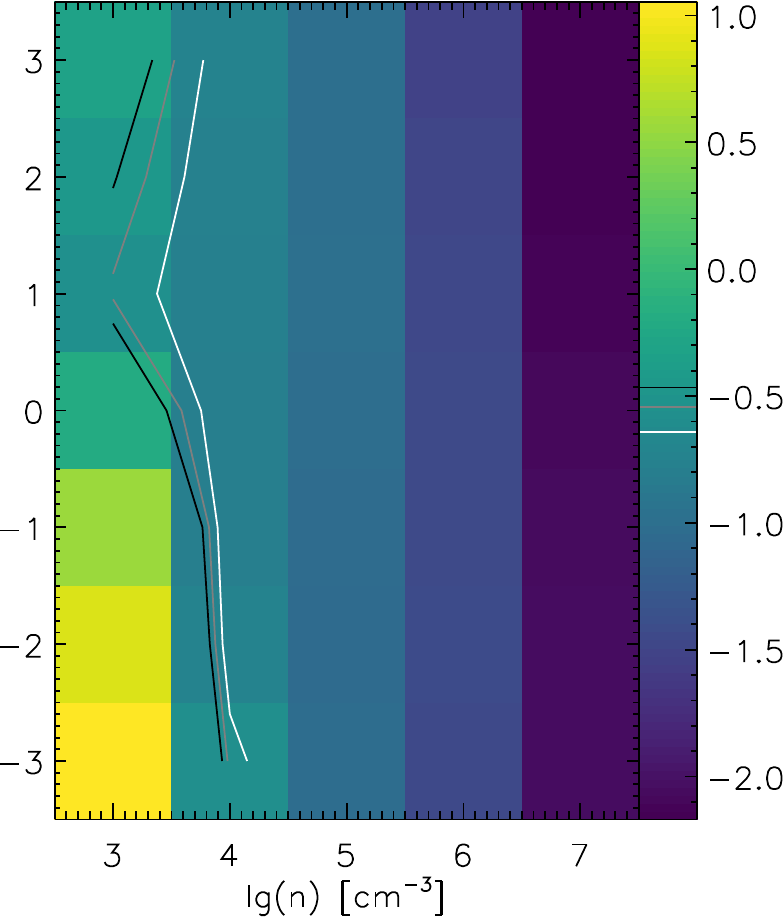}
   \includegraphics[angle=0,height=4.1cm]{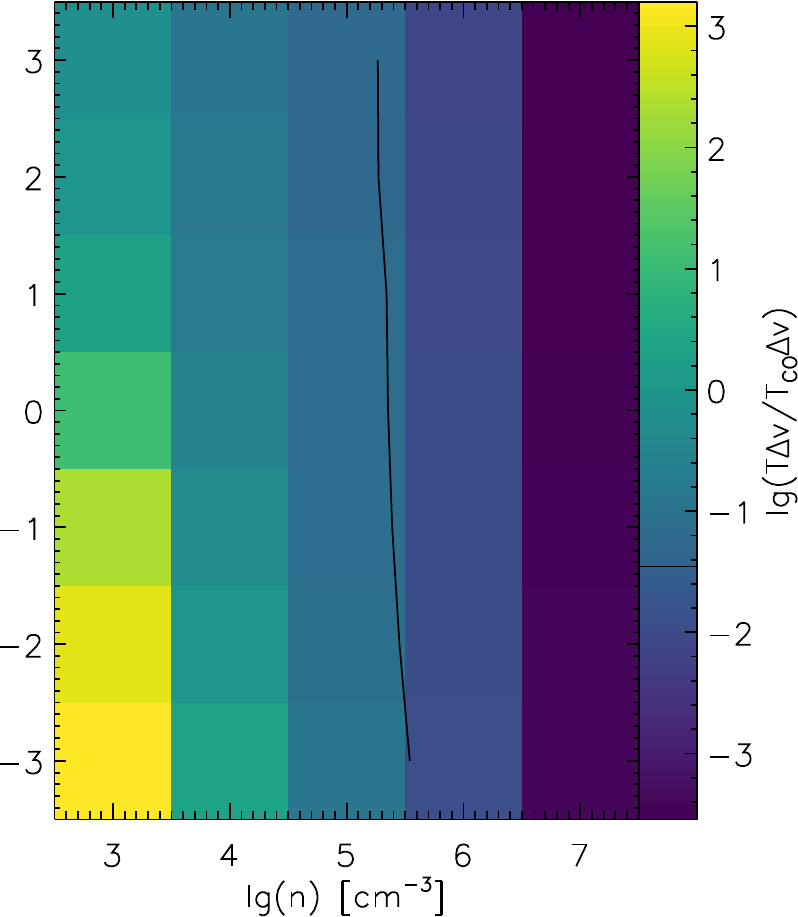}
   \caption{Same as Fig.~\ref{fig_model_relative_to_co} but for a UV radiation field of 10$\chi_0$.}
    \label{fig_model_relative_to_co_chi10}%
\end{figure*}
\FloatBarrier{}

\section{Models for an elevated radiation field}
\label{appx_pdr}

As there are no bright sources known in the vicinity of the considered cloud we expect that the standard UV field in the solar neighborhood also applies there. However, as the GAIA data are incomplete for embedded massive stars, there is a chance to find an elevated UV field, even if the Herschel continuum data give no hint for hotter dust. To be on the safe side, we also checked PDR models with a ten times higher UV flux in terms of the fit of the observational data.
Figures~\ref{fig_model_relative_to_nh_chi10} and \ref{fig_model_relative_to_co_chi10} show the intensity ratios from the model grid compared to the observations in the same way as Figs.~\ref{fig_model_relative_to_nh} and \ref{fig_model_relative_to_co}. We see qualitatively the same behavior. With the higher UV flux, the low observed intensities of $^{13}$CO 2-1 and C$^{18}$O 2-1 are met be the smallest clumps with low densities, but the non-detection of the \CII{} line is only met for models with very high densities and masses. The fundamental incompatibility of both observational constraints with the PDR modelling remains.

\end{appendix}
\end{document}